\documentclass[aps,prd,showpacs,superscriptaddress,nofootinbib]{revtex4-2}

\usepackage{graphicx}  
\usepackage{dcolumn}   
\usepackage{bm}        
\usepackage{amssymb,amsfonts,amsmath,physics}   
\usepackage{cancel}
\usepackage{mathtools}
\usepackage{slashed}

\usepackage{tikz-feynman}
\usepackage{stmaryrd}
\tikzfeynmanset{compat=1.1.0}
\usetikzlibrary{arrows}
\tikzset{	
	vertex/.style={circle,draw, minimum size=1.5em},	
	edge/.style={->,> = latex'}	
}
\usepackage[bookmarks, breaklinks, colorlinks,urlcolor=blue, citecolor=red, 
linkcolor=blue]{hyperref}
\usepackage[normalem]{ulem}

 \def\su{SU(2)_{\rm L} }
\def\L{{\rm L}}
\def\vp{\varphi}
\def\er{\eta_R}
\def\ei{\eta_I}
\def\l{\lambda}

\begin{document}
\rightline{CERN-TH-2026-218}
\title{Spontaneous Scoto-leptogenesis}

\author{Arghyajit Datta}
\email{arghyad053@gmail.com}
\affiliation{Department of Physics, Chung-Ang University, Seoul 06974, Korea}

\author{Hyun Min Lee}
\email{hminlee@cau.ac.kr}
\affiliation{Department of Physics, Chung-Ang University, Seoul 06974, Korea}
\affiliation{Theoretical Physics Department, CERN, 1211 Geneva, Switzerland}

\author{Jun-Ho Song}
\email{thdwnsgh1003@gmail.com}
\affiliation{Department of Physics, Chung-Ang University, Seoul 06974, Korea}

\begin{abstract} 
We propose a low-scale spontaneous leptogenesis scenario within the dynamical minimal scotogenic model for accommodating neutrino masses and inert scalar dark matter simultaneously. Thus, we dub the mechanism Spontaneous Scoto-leptogenesis. In this setup, a rolling Majoron arising from the global $U(1)_{B-L}$ symmetry breaking induces an effective chemical potential for the $B-L$ charge in the presence of $B-L$ violating interactions that allow for the efficient decays and inverse decays of right handed neutrinos (RHN) and RHN mediated scattering, so it gives rise to the observed baryon asymmetry of the Universe through the electroweak sphaleron conversion. The mechanism becomes effective in the strong washout regime and successfully lowers the viable mass scale of the lightest RHN to the range of TeV scales,  thereby making the thermal scotogenic leptogenesis with two hierarchical RHNs accessible to direct tests. 
We identify the roles of the $\l_5$ coupling for spontaneous leptogenesis and inert scalar dark matter through the efficient erasure of the inert scalar asymmetry.  We also explore the regime for Majoron dark matter from the kinetic misalignment, showing that a multicomponent dark sector comprising the inert scalar and  the Majoron can be realized in the model. 
The resulting framework provides a unified origin for low-scale baryogenesis, neutrino masses, and multicomponent dark sector, so it can be tested by complementary experimental probes through direct detection experiments, collider searches for inert scalars, and future detection of Majoron dark matter or dark radiation.
\end{abstract}
\maketitle

\section{Introduction}
One of the most compelling pieces of evidence for physics beyond the Standard Model (BSM) is the existence of  baryon asymmetry of the Universe (BAU) observed from the Cosmic Microwave Background (CMB) anisotropies~\cite{Planck:2018vyg}  and the light elemental abundances generated during Big Bang nucleosynthesis (BBN)~\cite{Mossa:2020gjc, Pisanti:2020efz, Yeh:2022heq}. The Standard Model (SM), despite incorporating baryon number violation through electroweak sphalerons, fails to reproduce the observed asymmetry because of its insufficient CP violation and the absence of a strongly first-order electroweak phase transition. This has motivated numerous mechanisms for dynamical generation of baryon asymmetry in the early Universe, among which leptogenesis provides one of the most compelling and well-motivated explanations by generating a primordial lepton asymmetry that is subsequently converted into a baryon asymmetry through electroweak sphaleron processes. 

In many conventional thermal leptogenesis scenarios~\cite{Fukugita:1986hr, Luty:1992un,  Barbieri:1999ma}, the generation of lepton asymmetry proceeds through the CP-violating decays of heavy right handed neutrinos (RHN), added on top of SM particles to explain nonzero SM neutrino masses~\cite{Minkowski_1977, Yanagida_1979, Yanagida_1979_1, GellMann_1979, Schechter_1980}. Although being remarkably successful, the viability of these setups for hierarchical RHN masses is often constrained by the interplay between CP asymmetry generation and washout processes, typically favoring a high RHN mass scale $M_1\gtrsim 10^{9}$ GeV that makes it hard to probe the RHN sector directly in experiments~\cite{Davidson:2002qv}. This limitation can be alleviated by considering a quasi-degenerate RHN mass spectrum, where the CP asymmetry is resonantly enhanced~\cite{Pilaftsis:1997jf,Pilaftsis:2003gt}. However, such a setup relies on an additional assumption about the RHN mass spectrum, whose origin typically requires further theoretical input, such as an underlying symmetry or otherwise necessitates an ad hoc fine-tuning of the RHN mass splitting. These considerations thus motivate the exploration of alternative leptogenesis mechanisms capable of generating the observed matter-antimatter asymmetry without relying on either a very high RHN mass scale or a finely structured RHN spectrum. 

An attractive alternative is the so called spontaneous baryogenesis~\cite{Cohen:1987vi, Cohen:1988kt}, which has been further developed for baryogenesis from leptogenesis~\cite{Li:2001st, Yamaguchi:2002vw, Kusenko:2014uta,Ibe:2015nfa,DeSimone:2016ofp, Bae:2018mlv,Co:2020jtv, Domcke:2020kcp, Co:2020xlh, Barnes:2022ren, Berbig:2023uzs, Chao:2023ojl, Chun:2023eqc, Barnes:2024jap, Wada:2024cbe, Datta:2024xhg, Chun:2024gvp, Chen:2025awt, Mishra:2025twb, Berbig:2025hlc, Chun:2025abp, Takahashi:2026ngu, Chun:2026jgn,Kuckenberg:2026oax}, baryogenesis from the QCD axion \cite{Co_2020,Chun:2025brc}, where the  rotation of a pseudo-Nambu-Goldstone boson(pNGB) induces an effective chemical potential for baryon (lepton) number. Then, while baryon (lepton)-number-violating interactions remain in thermal equilibrium, a net baryon (lepton) asymmetry is generated dynamically.  For the case of lepton asymmetry generation, once these lepton number violating interactions decouple, the generated asymmetry is preserved and subsequently converted into the observed baryon asymmetry through electroweak sphaleron processes~\cite{Harvey:1990qw}. Compared with the conventional thermal setup, the presence of the background chemical potential in this case qualitatively enhances the asymmetry generation. This opens up new regions of parameter space for successful leptogenesis. Such a nonzero chemical potential is generated when the radial partner of the pNGB is once far away from the minimum of the potential such that there is a kick along the direction of the pNGB due to an explicit breaking term of the global symmetry. For instance, the Hubble-induced mass for the radial partner could trap the radial partner in a local minimum during inflation \cite{Co:2019jts,Co_2020} or the radial partner can be the inflaton, kicking the pNGB to a nonzero rotation at the end of inflation \cite{Lee:2023dtw,Lee:2024bij,Chun:2025brc}.

A particularly appealing framework in which a similar idea can be realized is the scotogenic model~\cite{Ma:2006km}, where the SM is extended by two or more hierarchical  RHNs and an inert Higgs doublet (IHD). In this framework, light neutrino masses are generated radiatively at one loop, while an exact discrete symmetry stabilizes the lightest odd particle, providing a viable dark matter (DM) candidate.  Furthermore, the dynamical realization of this setup  provides a natural origin for the heavy mass scales by relating them to the spontaneous breaking of an underlying global symmetry~\cite{DeRomeri:2022cem, Portillo-Sanchez:2023kbz, Chun:2023vbh}. The corresponding symmetry-breaking scalar field simultaneously generates the masses of the RHNs and the dark-sector particles through its vacuum expectation value ($vev$), resulting in a more unified and theoretically well-motivated framework than scenarios in which these masses are introduced by hand.

The same framework also naturally accommodates leptogenesis through the out-of-equilibrium CP-violating decays of the heavy RHNs~\cite{Kashiwase:2012xd, Kashiwase:2013uy, Racker:2013lua, Clarke:2015hta, Hugle:2018qbw, Mahanta:2019gfe}. However, in conventional thermal leptogenesis, the setup remains subject to the tension between efficient asymmetry generation and strong washout effects. More specifically, in the minimal scotogenic model with two hierarchical RHNs, such a tension is translated into a lower bound of $M_1 \gtrsim 10^{11}$ GeV for successful leptogenesis~\cite{Hugle:2018qbw}, closely paralleling the type-I seesaw case. Relaxing this bound typically requires either a quasi-degenerate RHN spectrum or an extension of the minimal particle content, for example, by introducing a third hierarchical RHN~\cite{Kashiwase:2012xd, Kashiwase:2013uy, Racker:2013lua, Clarke:2015hta, Hugle:2018qbw}. Such possibilities, however, require additional structure beyond the minimal setup with two RHNs.

Interestingly, in the dynamical realization of the scotogenic model, the spontaneous breaking of the underlying global symmetry predicts a Majoron as a pseudo-Nambu–Goldstone boson.  The cosmological evolution of the Majoron provides precisely the additional ingredient required for spontaneous leptogenesis. Moreover, the coherent motion of the Majoron may also contribute to the present DM abundance through conventional~\cite{Preskill:1982cy, Abbott:1982af,Dine:1982ah} or kinetic~\cite{Co:2019jts} misalignment.
This opens up an interesting possibility where the same dynamics responsible for the origin of neutrino masses can also alleviate the limitations of conventional thermal leptogenesis and support a multicomponent dark sector consisting of the inert scalar and the Majoron.
In general, radiative neutrino mass, low-scale leptogenesis, and multicomponent DM can be connected with the minimal field content in the dynamical scotogenic scenario. To the best of our knowledge, such a realization has not been explored previously.

Motivated by this consideration, in this work, we investigate spontaneous leptogenesis in the dynamical minimal scotogenic model, where the spontaneous breaking of a global $U(1)_{B-L}$ symmetry gives rise to a Majoron field, $\theta(x)\equiv J(x)/v_\varphi$.  The time evolution of this Majoron background provides an effective chemical potential for $B-L$ charge through the shift-symmetric derivative interaction $\partial_\mu\theta\, j^\mu_{B-L}$ with fermions, while the decay and inverse decay of heavy Majorana RHN into leptons and the IHD and RHN mediated scattering provide the necessary $B-L$ violating interactions. The effective $B-L$ asymmetry then keeps on being generated till the inverse decays decouple as long as the Majoron keeps rolling.
Subsequently, the generated $B-L$ asymmetry is converted into the observed baryon asymmetry by electroweak sphalerons.

We remark that this way of generating baryon asymmetry is fundamentally different  from the conventional thermal scenario. Instead of generating the asymmetry through CP-violating RHN decays, a rolling CP-odd Majoron  here generates an effective $B-L$ chemical potential, while RHN-mediated $B-L$ violating interactions convert this dynamical bias into a net $B-L$ asymmetry. As a consequence, the tension between efficient asymmetry generation and washout is significantly alleviated, opening previously inaccessible regions of parameter space, including the low-mass RHN regime.  In particular, we demonstrate that the observed baryon asymmetry can be successfully generated with RHN masses as low as $M_1= \mathcal{O}(600)$ GeV while simultaneously accounting for radiative neutrino masses and the observed DM abundance, in sharp contrast to the conventional thermal leptogenesis scenario. Our framework, therefore, provides a unified, economical and testable explanation for the origin of the observed BAU, neutrino masses, and DM.

The remainder of this paper is organized in the following manner. In Sec.~\ref{sec:neutrino} we briefly review the process of neutrino mass generation in the minimal scotogenic model and summarize its key features. In Sec.~\ref{sec:dm},  we discuss the
IHD DM phenomenology and collect the neutrino mass and DM constraints relevant for spontaneous leptogenesis.  In Sec.~\ref{sec:lepto1}, we first discuss the limitations of conventional thermal leptogenesis in the minimal scotogenic model and then present a detailed analysis of the spontaneous leptogenesis framework with two hierarchical RHNs in Sec.~\ref{sec:lepto2}. The role of Majoron condensate as dominant DM is also explored in this section. We summarize our results in Sec.~\ref{sec:lepto3} and finally conclude in Sec.~\ref{sec:con}.

\section{Dynamical scotogenic model and neutrino masses}\label{sec:neutrino}

\begin{table}[h]
	\centering
	\begin{tabular}{|c|c|c|c|c|c|c|}
		\hline \phantom{XXXXXXXX} & \phantom{X} $\ell_{L}$ \phantom{X} & \phantom{X} $\Phi$ \phantom{X} & \phantom{X} $\eta$ \phantom{X} & \phantom{X} $N$ \phantom{X} & \phantom{X} $\varphi$ \phantom{X} \\
		\hline \hline
		  $\su \otimes U(1)_Y$ & {\bf  $(2,-1/2)$}  & {\bf  $(2,1/2)$} & {\bf  $(2,1/2)$} & {\bf  $(1,0)$} & {\bf  $(1,0)$} \\
		\hline
		{ $U(1)_{B-L}$} & {$-1$} &  {$0$} & {$0$} & {$-1$} & {$2$}  \\
		\hline
		$\mathbb{Z}_2$ & $+$  & $+$  & $-$ & $-$ & $+$ \\
		\hline
	\end{tabular}
	\caption{\centering Particle content and charge assignments of Scotogenic model. \label{tab:particle-content}}
\end{table}
The scotogenic model proposed by Ma~\cite{Ma:2006km} contains one additional SM doublet scalar and two or more RHNs to explain the nonzero neutrino masses. In our work, we extend the scotogenic model with an additional complex scalar field $\vp$, which is also an SM singlet. With an appropriate choice of leptonic and $\mathbb{Z}_2$ charge for $\vp$, this scenario leads to a dynamical scotogenic extension, which respects additional global $U(1)_{B-L}$ symmetry on top of SM gauge and  $\mathbb{Z}_2$ symmetry at high temperature.  In that case, we have the $SU(3)_c\otimes SU(2)_\L\otimes U(1)_Y\otimes U(1)_{B-L}\otimes \mathbb{Z}_2$ symmetry as summarized in Table~\ref{tab:particle-content} and the symmetry-invariant Yukawa Lagrangian relevant for neutrino masses is given by
\begin{equation}
	-\mathcal{L}\supset Y^\nu_{\alpha i} \bar{\ell}_{L_\alpha} \tilde{\eta} N_i + \frac12 Y^N_{i} \varphi \bar{N^c_i} N_i  +{\rm h.c.},
	\label{eq:LagYL}
\end{equation}
where $\tilde{\eta}\equiv\sigma_2\eta^*$ is the conjugate scalar doublet and $\ell_\L^\alpha\equiv(\nu_{\L},\alpha_\L)^T$ (with $\alpha=e,~\mu,~\tau$) are the SM lepton doublets. The minimal scotogenic model is defined with two RHNs for which $i=1,2$. The scalar potential complying with the symmetry takes the form
\begin{widetext}
	\begin{align}
		V &=  \mu_\Phi^2 \Phi^\dagger \Phi + m_\eta^2 \eta^\dagger \eta + m_\vp^2 \vp^* \vp + \frac{\lambda_\Phi}{2} (\Phi^\dagger \Phi)^2 + \frac{\lambda_\eta}{2} (\eta^\dagger \eta)^2 + \lambda_3 (\eta^\dagger \eta)(\Phi^\dagger \Phi) + \lambda_4 (\eta^\dagger \Phi) (\Phi^\dagger \eta ) \nonumber \\
		&+ \displaystyle \frac{\lambda_5}{2} \left[ (\eta^\dagger \Phi)^2 + \text{h.c} \right] + \frac{\lambda_\vp}{2} (\vp^* \vp)^2 + \lambda_{\Phi\vp} (\Phi^\dagger \Phi)(\vp^* \vp) + \lambda_{\eta\vp} (\eta^\dagger \eta) (\vp^* \vp),
		\label{eq:potential-dynamical}
	\end{align}
\end{widetext}
where all the parameters in the potential can be chosen to be real without loosing the generality. 

The spontaneous breaking of the global $U(1)_{B-L}$ in this case is triggered by the nonzero $vev$ $v_\vp$ of the $\vp$, which subsequently generates the Majorana mass for the RHNs $M_i= Y^N_i v_\vp/\sqrt{2}$ while the $vev$ of $\Phi$ drives the electroweak symmetry breaking (EWSB). Note that the conservation of the $\mathbb{Z}_2$ symmetry prevents a nonzero $vev$ for $\eta$, which also ensures one of the components of $\eta$ to be a viable DM candidate. 

After EWSB, the scalar fields can be identified as
$ \Phi=
\left(
0,
(v_\Phi+h_\Phi)/\sqrt{2}\right)^T
$, $
\eta=
\left( \eta^+,
(\er+i\ei)/\sqrt{2}\right)^T,
$ 
and  
$ 
\varphi=\frac{1}{\sqrt{2}}\left(v_\vp+h_\vp\right)
\exp\left[i\frac{J(x)}{v_\vp}\right],
$ where $\theta\equiv J/v_\vp$ is the Majoron field.
Consequently, the mass of the components of $\eta$ field takes the form
\begin{align}
	m_{\er}^2&= m_\eta^2 +\frac{v_\Phi^2}{2}(\lambda_3+\lambda_4+\lambda_5) +\frac{v_\vp^2}{2} \lambda_{\eta \vp},\notag\\
	m_{\ei}^2&= m_\eta^2 +\frac{v_\Phi^2}{2}(\lambda_3+\lambda_4-\lambda_5)+\frac{v_\vp^2}{2} \lambda_{\eta \vp},\\
	m_{\eta_\pm}^2&= m_\eta^2 +\frac{v_\Phi^2}{2}\lambda_3 +\frac{v_\vp^2}{2} \lambda_{\eta \vp}.
	\notag
\end{align}
The mass matrix for the other scalars can be extracted from the Lagrangian
\begin{align}
	\mathcal{L}\supset-\frac12 \begin{pmatrix} h_\Phi & h_\vp\end{pmatrix}
	\begin{pmatrix}
		\lambda_\Phi v_\Phi^2 & \lambda_{\Phi \vp} v_\Phi v_\vp\\
		\lambda_{\Phi \vp} v_\Phi v_\vp & \lambda_\vp v_\vp^2
	\end{pmatrix}
	\begin{pmatrix} h_\Phi \\ h_\vp\end{pmatrix},
\end{align}
diagonalization of which leads to the  mass of the SM Higgs $h$ and a heavier scalar $\phi$, given by
\begin{align}
	m_{\phi, h}^2= \frac12\left( \lambda_\Phi v_\Phi^2 +  \lambda_\vp v_\vp^2 \pm
	\sqrt{\left( \lambda_\Phi v_\Phi^2 - \lambda_\vp v_\vp^2\right)^2 +4 \lambda_{\Phi \vp} v_\Phi^2 v_\vp^2}
	\right).
\end{align}
In this case, 
notice that the stability of the  lightest component of the $\eta$ particle is confirmed due to the presence of the exact $\mathbb{Z}_2$ symmetry.

Given the field content, the SM neutrino mass is generated when both $U(1)_{B-L}$ and SM gauge symmetry break down. Interestingly, the presence of exact $\mathbb{Z}_2$ symmetry although forbids any tree-level terms for neutrino mass, at 1-loop level, it can be generated when RHNs and the inert scalars $\eta_{R,I}$ circulate in the loop. The neutrino mass matrix then takes the form~\cite{Ma:2006km}
\begin{widetext}
	\begin{align}
		&\left(M_\nu\right)_{\alpha \beta}= \sum_{i} \frac{M_i Y^\nu_{\alpha i} Y^\nu_{\beta i}}{32 \pi^2}
		\left[L^{(i)}_{\er}-L^{(i)}_{\ei}\right]=\frac{v_{\Phi}^2}{2} \sum_i Y^\nu_{\alpha i} \left( \frac{1}{M_i} \xi_i \right) [(Y^{\nu})^T]_{i\beta},
		\label{eq:scotogenic-mass}
	\end{align}
\end{widetext}
with
\begin{align}
	\xi_i
	\equiv
	\frac{M_i^2}{16\pi^2v_{\Phi}^2}
	\left[L^{(i)}_{\er}-L^{(i)}_{\ei}\right], ~~L^{(i)}_x= \frac{m_{x}^2}{m_{x}^2-M_i^2}\ln\left(\frac{m_{x}^2}{M_i^2}\right).
\end{align}
Notice that the third equality in Eq.~\eqref{eq:scotogenic-mass} is analogous to the conventional type-I seesaw formula, with the inverse RHN mass $M_i^{-1}$ effectively replaced by $\xi_i/M_i$, where the dimensionless factor $\xi_i$ encodes the loop-induced suppression as well as the dependence on the IHD mass splitting, governed by the coupling $\l_5$.  Interestingly, at an energy $E\ll M_i$, when RHNs are already integrated out,  $\l_5\to 0$ leads to degenerate masses for $\eta_{R}$ and $\ei$. Consequently,  SM neutrino mass evaluated in Eq.~\eqref{eq:scotogenic-mass} vanishes. This reflects the restoration of an accidental global symmetry of the scalar sector in the $\l_5\to0$ limit, under which the inert doublet carries a conserved quantum number. Therefore, the smallness of neutrino masses is naturally associated with the smallness of the symmetry-breaking parameter $\l_5$ \cite{Kim:2025decoupled,Kim:2025zrs}.

The neutrino mass matrix $M_{\nu}$ here is a complex symmetric matrix, which is diagonalized by the Pontecorvo-Maki-Nakagawa-Sakata (PMNS) mixing matrix $U$ as $D_m= U^T M_\nu U$~\cite{Pontecorvo:1957cp,Maki:1962mu}. As a result, using  Casas-Ibarra (CI) parametrization ~\cite{Casas_2001, Hugle:2018qbw}, the Yukawa couplings can be expressed as 
\begin{align}
	Y^\nu= U^* D_{\sqrt{m}} R D_{\sqrt{\Lambda}},
	\label{eq:ci}
\end{align}
where $R$ is a complex orthogonal matrix which satisfy $R^T R=1$ and $D_{\sqrt{\Lambda}}= \text{diag}[\Lambda_{ii}^{1/2}]$ with $\Lambda_{ii}= \frac{2M_i}{v_\Phi^2} \xi_i^{-1}$. Notice that while constructing the Yukawa structure in Eq.~\eqref{eq:ci} we have implicitly assumed that the CP odd part of IHD, i.e., $\ei$ is the only DM candidate in the IHD sector, which can be realized for a positive value of $\lambda_5$ parameter. In that case, $\Lambda_{ii}$ will also be positive.

In this dynamical scotogenic setup,  extending the fermionic sector with only two RHNs would be sufficient to explain the SM neutrino masses. In that case, normal ordered light neutrino masses ($m_1<m_2<m_3$) satisfy
\begin{align}
	D_m= \text{diag}[0,\sqrt{\Delta m_{21}^2},\sqrt{\Delta m_{31}^2}],
	\label{eq:lightmass}
\end{align}  
where $\Delta m_{21}^2$ ($\Delta m_{31}^2$) is the observed solar (atmospheric) neutrino mass squared difference.  Making a further assumption of $M_1<M_2$ but keeping two RHNs contributing to the neutrino mixings, we would have the simplest structure of $R$ matrix as~\cite{Hugle:2018qbw}
\begin{align}
	R=\begin{pmatrix}
		0 & 0\\
		\cos\vartheta & -\sin\vartheta\\
		\sin\vartheta & \cos\vartheta
	\end{pmatrix},~~ 
	\text{with}~\vartheta= \vartheta_{r}+ i \vartheta_{i}. \label{Rmatrix}
\end{align}
As a result, for the observed SM neutrino mass differences and mixings (fixed from the global fitting of neutrino oscillation observations~\cite{Esteban:2024eli}), the exact form of the Yukawa matrix $Y^\nu$ can be generated by knowing the parameters $M_i,~m_{\ei},~\l_5$ and $\vartheta$, defined in Eq.~(\ref{Rmatrix}). Since $\ei$ plays the role of one of the DM candidates in this setup (apart from the Majoron),  the parameter $m_{\ei}$ and $\l_i(i=3,~4,~5)$ can be fixed by knowing the exact relic contribution of $\ei$ to DM abundance. So, for a fixed value of $M_i$, parameter $\vartheta$ controls the entries of $Y^\nu$.

In the subsequent sections, we will discuss how the DM parameters are fixed, followed by the discussion about the dependence of $Y^\nu$ on baryon asymmetry generation via thermal leptogenesis from RHN decay, the limitation of the standard thermal leptogenesis mechanism, and the natural emergence of spontaneous leptogenesis within this setup.

\section{Inert doublet dark matter}
\label{sec:dm}

The dark matter phenomenology with the scotogenic model has been extensively studied in the literature~\cite{LopezHonorez:2006gr,Hambye:2009pw, Dolle:2009fn,Arina:2009um, LopezHonorez:2010tb, LopezHonorez:2010qf,Goudelis:2013uca,Krawczyk:2013jta, Diaz:2015pyv, Garcia-Cely:2015khw, Belyaev:2016lok, Eiteneuer:2017hoh, Choubey:2017hsq, Borah:2017dfn, Kalinowski:2018ylg, Borah:2019aeq, Jueid:2020qis, Bose:2024bnp,  Kim:2025decoupled,Ardila-Tafurth:2025qdf,Lee:2026global}. In this section we summarize the aspects of those analyses that are relevant for the present work, with the main goal of identifying the allowed range of dark-sector parameters entering spontaneous leptogenesis after imposing the constraints from DM and neutrino mass phenomenology.

The exact $\mathbb{Z}_2$ symmetry introduced in the scotogenic model makes the lightest $\mathbb{Z}_2$-odd particle stable. The electrically charged inert scalars (although $\mathbb{Z}_2$ odd) cannot be the DM. Thus, in the present particle content this leaves three qualitatively different possibilities: the lightest RHN $N_1$, the CP-even neutral scalar $\er$, or the CP-odd neutral scalar $\ei$ (apart from Majoron condensate, which we will discuss in the later sections). Note that, in this work, the RHNs are required to participate in the lepton-number-violating reactions that generate the baryon asymmetry; in particular, the lightest state must be able to decay through $N_1\to \ell \eta$ and the corresponding inverse process. We therefore take the stable relic to be the lightest neutral component of the inert doublet.

In what follows we choose the CP-odd scalar $\ei$ as the DM particle in IHD sector. This choice is not a loss of generality for the phenomenological discussion, since the CP-even case is obtained by the interchange $\ei\leftrightarrow\er$ together with the corresponding sign choice for $\lambda_5$. With our convention in Eq.~\eqref{eq:potential-dynamical}, $\ei$ is lighter than $\er$ for $\l_5>0$,
while the requirement that the charged scalar is heavier than the DM state gives
$m_{\eta_\pm}>m_{\ei}.$ It is useful to trade the quartic coupling $\lambda_4$ for the coupling that directly controls the Higgs portal interaction of the DM field,
$ \lambda_{\rm DM}\equiv \lambda_3+\lambda_4-\lambda_5 .
$
The four parameters that will be varied in the DM analysis are therefore
\begin{align}
	\left\{m_{\ei},~\lambda_{\rm DM},~\lambda_3,~\lambda_5\right\}.
	\label{eq:dm-free-parameters}
\end{align}
They determine the overall inert doublet mass scale, the strength of the Higgs-mediated annihilation and direct detection (DD) interactions, and the mass splittings among the three inert components.

For the analysis we work in the limit where the singlet scalar does not affect the inert doublet freeze-out dynamics. More explicitly, we take the portal couplings connecting the singlet to the SM and inert doublets to be negligible,
$
\lambda_{\Phi\varphi}\simeq 0,~
\lambda_{\eta\varphi}\simeq 0,
$
so that the scalar $h_\Phi$ is identified with the observed Higgs boson and
$ \lambda_\Phi=\frac{m_h^2}{v_\Phi^2}$.
We also fix $\lambda_\eta=0.2$, which is safely perturbative and keeps the inert self-interaction away from the region where tree-level unitarity can become restrictive. In terms of the parameters in Eq.~\eqref{eq:dm-free-parameters}, the inert doublet mass parameters then take the form
\begin{align}
	m_\eta^2 &= m_{\ei}^2-\frac12\lambda_{\rm DM}v_\Phi^2,\notag\\
	m_{\er}^2 &= m_{\ei}^2+\lambda_5 v_\Phi^2,\label{eq:inert-mass-free}\\
	m_{\eta_\pm}^2 &= m_{\ei}^2+\frac12\left(\lambda_3-\lambda_{\rm DM}\right)v_\Phi^2.\notag
\end{align}
Thus the condition that $\ei$ is the lightest inert state becomes
\begin{align}
	\lambda_5>0,\qquad \lambda_3>\lambda_{\rm DM}.
	\label{eq:etaI-lightest}
\end{align}
The neutral mass splitting is particularly important phenomenologically. A nonzero $\lambda_5$ splits $\ei$ and $\er$, kinematically suppresses the dangerous tree-level inelastic coupling of the DM state to the $Z$ boson~\cite{Arina:2009um,LopezHonorez:2010tb}, and at the same time enters the one-loop neutrino mass~\cite{Ma:2006km} in Eq.~\eqref{eq:scotogenic-mass}. Consequently, the same parameter that distinguishes the two neutral inert scalars also controls the size of the radiative neutrino mass for fixed Yukawa couplings and RHN masses.

The quartic couplings in this setup are also not arbitrary. They determine whether at large field values, the scalar potential is bounded from below~\cite{Deshpande:1977rw,Swiezewska:2012eh,Branchina:2024uhl}, so the vacuum-stability conditions for the quartic couplings are given as follows,
\begin{align}
	&\lambda_\Phi>0,\qquad
	\lambda_\eta>0,\qquad \lambda_\varphi>0, 
	\\
	&\lambda_3+\lambda_4+\lambda_5+\sqrt{\lambda_\Phi\lambda_\eta}>0, \qquad \lambda_{\Phi\varphi}+\sqrt{\lambda_\Phi \lambda_\varphi}>0,\qquad \lambda_{\eta\varphi}+\sqrt{\lambda_\eta \lambda_\varphi}>0, \\
	&\sqrt{\lambda_\Phi \lambda_\eta \lambda_\varphi}+(\lambda_3+\lambda_4+\lambda_5)\sqrt{\lambda_\varphi} +\lambda_{\Phi\varphi} \sqrt{\lambda_\eta}+\lambda_{\eta\varphi}\sqrt{\lambda_\Phi} \\
	&+\frac{1}{2}\sqrt{(\lambda_3+\sqrt{\lambda_\Phi \lambda_\eta})(\lambda_{\Phi\varphi}+\sqrt{\lambda_\Phi\lambda_\varphi})(\lambda_{\eta\varphi}+\sqrt{\lambda_\eta\lambda_\varphi})}>0.
	\label{eq:vacuum-stability-dm}
\end{align}
For insance, focusing on the bounds on $\lambda_{3,4,5}$, we can use $m_h= 125~{\rm GeV}$, $v_\Phi= 246~{\rm GeV}$ and  a benchmark value $\lambda_\eta=0.2$ to obtain
\begin{align}
	\lambda_3> -0.16,\qquad
	\lambda_{\rm DM}+2\lambda_5> -0.16.
\end{align}
The other vacuum stability conditions can be also satisfied for appropriate values of the other quartic couplings.
Moreover, the quartic couplings in the scalar potential are subject to the perturbativity conditions \cite{Branco:2011iw,Bian:2017xzg}, as follows,
\begin{align}
&\lambda_\Phi < 8\pi, \qquad \lambda_\eta<8\pi, \qquad \lambda_\varphi<8\pi, \\
&|\lambda_3\pm \lambda_4|<8\pi, \qquad |\lambda_3+2\lambda_4\pm 3\lambda_5|<8\pi, \qquad |\lambda_3\pm \lambda_5|<8\pi, \\
&\frac{1}{2}(\lambda_\Phi+\lambda_\eta) \pm \frac{1}{2}\sqrt{(\lambda_\Phi-\lambda_\eta)^2+4\lambda^2_4}<8\pi, \\
&\frac{1}{2}(\lambda_\Phi+\lambda_\eta) \pm \frac{1}{2}\sqrt{(\lambda_\Phi-\lambda_\eta)^2+4\lambda^2_5}<8\pi, \\
&\frac{3}{2}(\lambda_\Phi+\lambda_\eta) \pm\sqrt{\frac{9}{4}(\lambda_\Phi-\lambda_\eta)^2+(2\lambda_3+\lambda_4)^2}<8\pi, \\
&|\lambda_{\Phi\varphi}|<8\pi, \qquad |\lambda_{\eta\varphi}|<8\pi.
\end{align}
In the subsequent discussions we will impose the above vacuum-stability requirements together with the perturbativity conditions for the quartic couplings.

Collider searches and electroweak precision measurements further constrain these parameters, especially in the regions with light inert scalars or sizable mass splittings within the inert doublet. The LEP measurement of the $Z$ width excludes an open decay channel $Z\to \ei\er$, which requires~\cite{Lundstrom:2008ai,Dolle:2009fn}
\begin{align}
	m_{\ei}+m_{\er}>m_Z.
\end{align}
Similarly, searches for charged scalar production and the measured $W$ width disfavor spectra in which $\eta^\pm$ is too light or where two-body decays such as $W^\pm\to \eta^\pm\eta_{I,R}$ are kinematically open. These constraints are most relevant when the inert states lie near or below the electroweak scale. Electroweak precision observables, especially the oblique parameter $T$, constrain large isospin breaking inside the inert doublet. In practice, this means that the mass gaps controlled by $\lambda_5$ and $(\lambda_3-\lambda_{\rm DM})$ combination cannot be taken arbitrarily large for a fixed $m_{\ei}$. In the high-mass regime, the bounds are milder, but they are still imposed by requiring the inert contribution to the oblique parameters to remain within the experimentally allowed region~\cite{Belyaev:2016lok}.

To avoid the strong constraints associated with on-shell electroweak gauge boson decays and LEP searches in the low-mass region of DM,  we will restrict the DM analysis to the heavy inert doublet regime, $m_{\ei}>400~{\rm GeV}$ in the subsequent sections. The viable explodable region used below can therefore be summarized as
\begin{align}
	m_{\ei}>400~{\rm GeV};~
	\lambda_5>0;~
	\lambda_3>\lambda_{\rm DM};~
	\lambda_3,\lambda_{\rm DM}\gtrsim -0.23,
	\label{eq:dm-scan-summary}
\end{align}
with $\lambda_\eta=0.2$ and $\lambda_{\eta\varphi}=\lambda_{\Phi\varphi}=0$. Additional upper limits on the quartic couplings are taken from perturbativity and tree-level unitarity, while the allowed mass splittings are checked against electroweak precision data.

\begin{figure}[t]
	\includegraphics[width=0.44\textwidth]{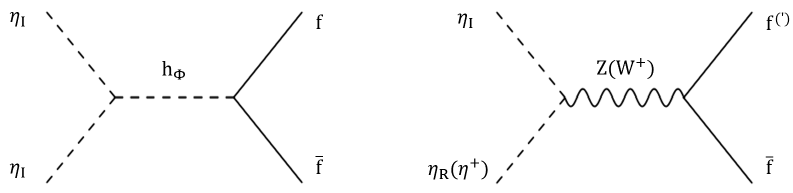} \,\,\,\,
	\includegraphics[width=0.49\textwidth]{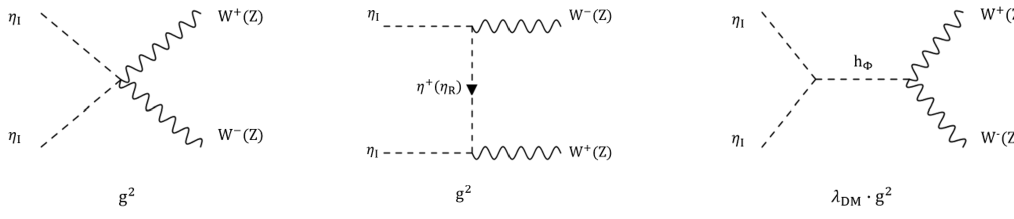} \vspace{0.2cm}\\
		\includegraphics[width=0.49\textwidth]{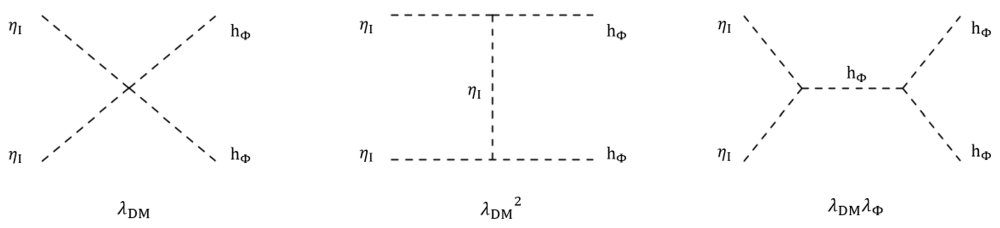} \,\,\,\,
	\includegraphics[width=0.34\textwidth]{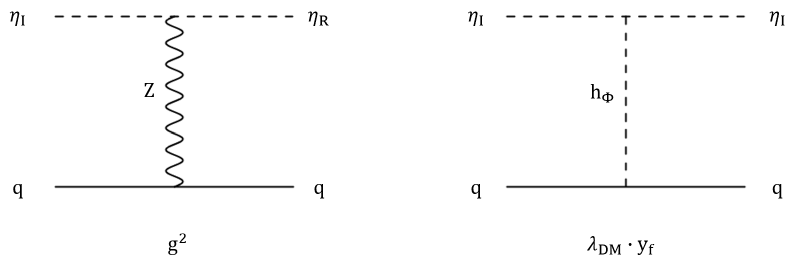}
	\caption{Feynman diagrams relevant for the DM annihilations and the DM direct detection. In the limit of vanishing Higgs mixings, $h_\Phi\simeq h$. }
	\label{fig:diagrams}
\end{figure}

The relic abundance of the DM can then be obtained by following the thermal evolution of the stable inert state. Since $\ei$, $\er$, and $\eta^\pm$ can be close in mass, coannihilations are generally important. The appropriate Boltzmann equation is therefore written in terms of the total number density of the coannihilating $\mathbb{Z}_2$-odd inert states,
\begin{align}
	\frac{d n}{dt}+3\mathcal{H}n=-\langle \sigma_{\rm eff} v\rangle
	\left(n^2-n_{\rm eq}^2\right),
	\label{eq:dm-boltzmann}
\end{align}
where $\mathcal{H}$ is the Hubble expansion rate and $\langle \sigma_{\rm eff} v\rangle$ includes annihilation and coannihilation channels into SM final states. The gauge interactions of the inert doublet dominate when the spectrum is compressed, while the Higgs portal coupling $\l_{\rm DM}$ controls the scalar-mediated contribution. After freeze-out, the present relic abundance is evaluated as
\begin{align}
	\Omega_{\ei} h^2=
	2.755\times 10^8
	\left(\frac{m_{\ei}}{{\rm GeV}}\right)
	\left(\frac{n_0}{s_0}\right),
	\label{eq:dm-relic}
\end{align}
where $n_0$ and $s_0$ denote the present DM number density and entropy density, respectively. The parameter points retained in the following analysis are those for which Eq.~\eqref{eq:dm-relic} reproduces the observed DM abundance within the cosmological uncertainty~\cite{Planck:2018vyg}. In Fig.~\ref{fig:diagrams}, we list the relevant Feynman diagrams for the DM annihilations and the DM direct detection.

\begin{figure}[t]
	\includegraphics[width=0.5\textwidth]{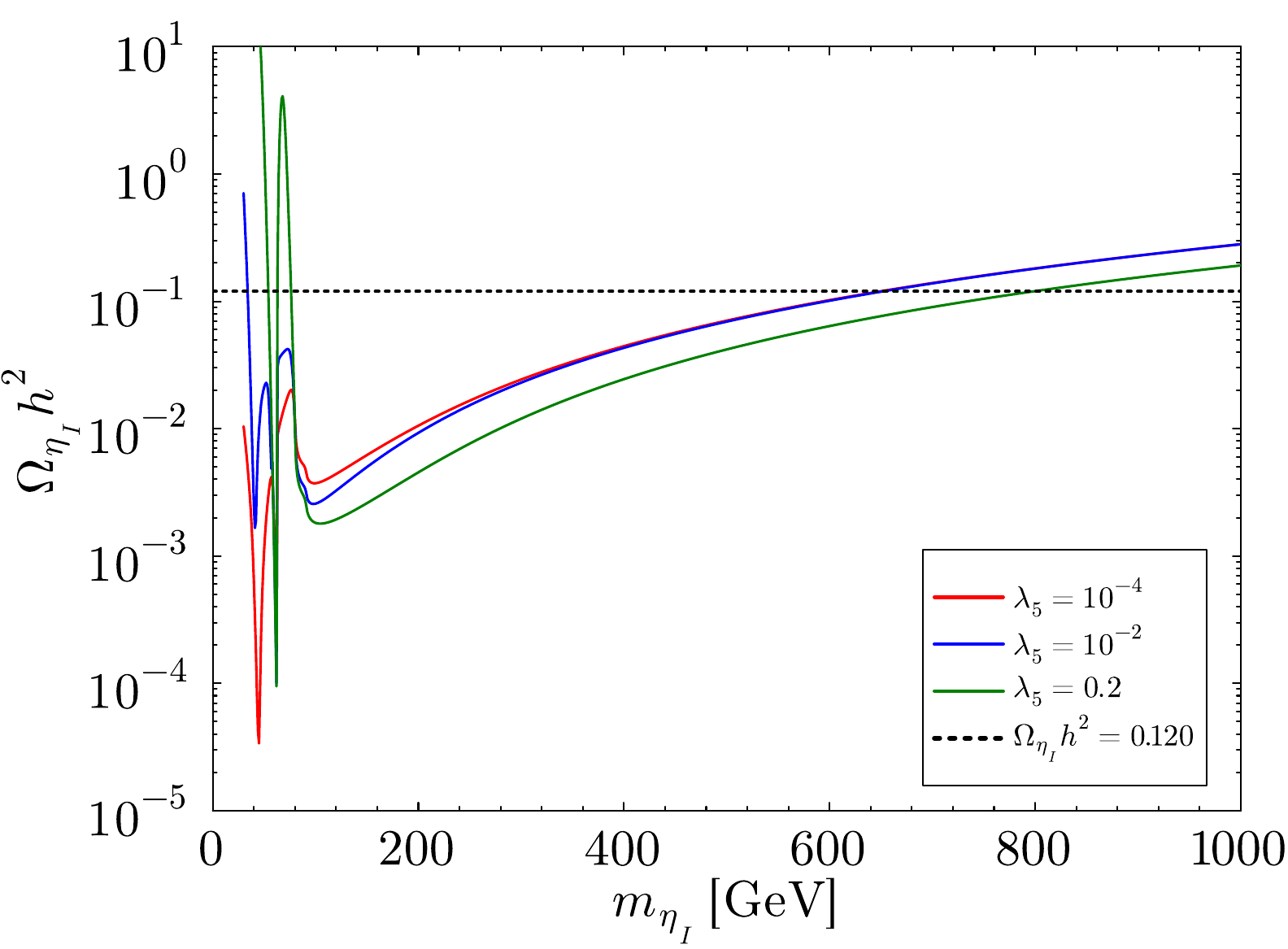}
	\caption{Relic density of the $\ei$ as a function of
		$m_{\ei}$ for fixed $\l_{\rm DM}=0.01$ and $\l_3=0.2$.
		The red, blue and green curves correspond to $\l_5=10^{-4}$,
		$10^{-2}$ and $0.2$ respectively. The horizontal dashed line indicates
		the observed value $\Omega_{\ei}h^2=0.120$.
		\label{fig:dmrelic}}
\end{figure}

Following the procedure outlined above, we compute the thermal relic abundance of the lightest inert scalar, $\ei$, using the public packages LanHEP~\cite{Semenov:2014rea} and \texttt{micrOMEGAs}~\cite{Alguero:2023gkk} for several benchmark values of $\l_5$. The variation of the relic density $\Omega_{\ei}$ with respect to the DM mass $m_{\ei}$ (for fixed value of $\l_{\rm DM}, \l_3$) is shown in Fig.~\ref{fig:dmrelic}.
The relic curves exhibit several characteristic features in the low DM mass region, caused by resonant annihilation and the onset of new annihilation channels.  For small values of $\l_5$, three prominent depletion regions appear at $m_{\ei}<400$ GeV. The first dip occurs at $m_{\ei}\sim 40-45$ GeV (around $m_{\ei}+m_{\er}\sim m_Z$), due to the resonance enhancement of $Z$-mediated coannihilation processes, $\ei \er \to Z^*\to \text{SM, SM}$, with ${\rm SM}=f, W,Z$, as shown in Fig.~\ref{fig:diagrams}. The second depletion is observed around $m_{\ei}\sim m_h/2$, where the Higgs-mediated annihilation channels, $\eta_I\eta_I\rightarrow h^{*}\rightarrow \mathrm{SM,SM}$, with ${\rm SM}=f,W,Z,h$, become resonant, as shown in Fig.~\ref{fig:diagrams}. A broader suppression around $m_{\eta_I}\sim 100$ GeV arises from the opening and increasing efficiency of electroweak gauge-boson final states, mainly $WW$ and $ZZ$, together with coannihilation channels involving $\eta_R$ and $\eta^\pm$.

For larger $\l_5$, $m_{\er}$ becomes much heavier. Consequently, the number density of $\er$ faces Boltzmann suppression at the time of freeze-out, and $\ei \er$ coannihilation resonance does not contribute significantly. This explains the disappearance of the low-mass coannihilation dip for larger $\l_5$, as illustrated by the green curve in Fig.~\ref{fig:dmrelic}. Consequently, for larger $\l_5$ one can get correct relic density for low $\ei$ mass regime $m_{\ei}<100$ GeV (although heavily constrained as discussed previously). In the heavy DM mass regime, i.e., $m_{\ei}>400$ GeV, a correct relic can again be achieved for $m_{\ei}\gtrsim 553$ GeV and with increasing $\l_5$ (i.e., increasing $\er$-$\ei$ mass splitting), the correct relic requirement shifts the mass toward larger values, which is also evident from the figure. 

In the region of $m_{\eta_I}\lesssim 100\,{\rm GeV}$, $\eta_I \eta_I\to WW,ZZ$ channels with gauge and quartic interactions are dominant for determining the relic density, and  the co-annihilation channels are also relevant for a small value of $\lambda_5$. But, for a fixed value of $\lambda_{\rm DM}=0.01$ in Fig.~\ref{fig:dmrelic},  for which $\eta_I \eta_I\to WW,ZZ$ are accordingly fixed, the co-annihilation channels reduce the relic density further for small values of $\lambda_5$ as shown for $\lambda_5=10^{-2}, 10^{-4}$. On the other hand, for $m_{\eta_I}\gg 100\,{\rm GeV}$,  according to the Goldstone equivalence theorem, we can focus on the DM annihilation channels with longitudinal components of gauge bosons and Higgs boson, $\eta_I\eta_I\to W_LW_L, Z_L Z_L,hh$, whose annihilation cross sections are proportional to  $\lambda^2_3$, $(\lambda_{\rm DM}+2\lambda_5)^2$ and $\lambda^2_{\rm DM}$, respectively. So, for $\lambda_5=10^{-4}-10^{-2}$, $\lambda_3=0.2$ and $\lambda_{\rm DM}=0.01$, we get $\lambda_{\rm DM}+2\lambda_5=0.01-0.03$, so $\eta_I \eta_I\to W_LW_L$ is dominant for $m_{\eta_I}\gg 100\,{\rm GeV}$. On the other hand, for $\lambda_5=0.2$, $\lambda_3=0.2$ and $\lambda_{\rm DM}=0.01$, we get  $\lambda_{\rm DM}+2\lambda_5=0.4$, so $\eta_I \eta_I\to Z_LZ_L$ contributes more importantly for $m_{\eta_I}\gg 100\,{\rm GeV}$. As a result, we find the reduction of the relic density for  $\lambda_5=0.2$ in Fig.~\ref{fig:dmrelic}.

\begin{figure}[t]
	\includegraphics[width=0.45\textwidth]{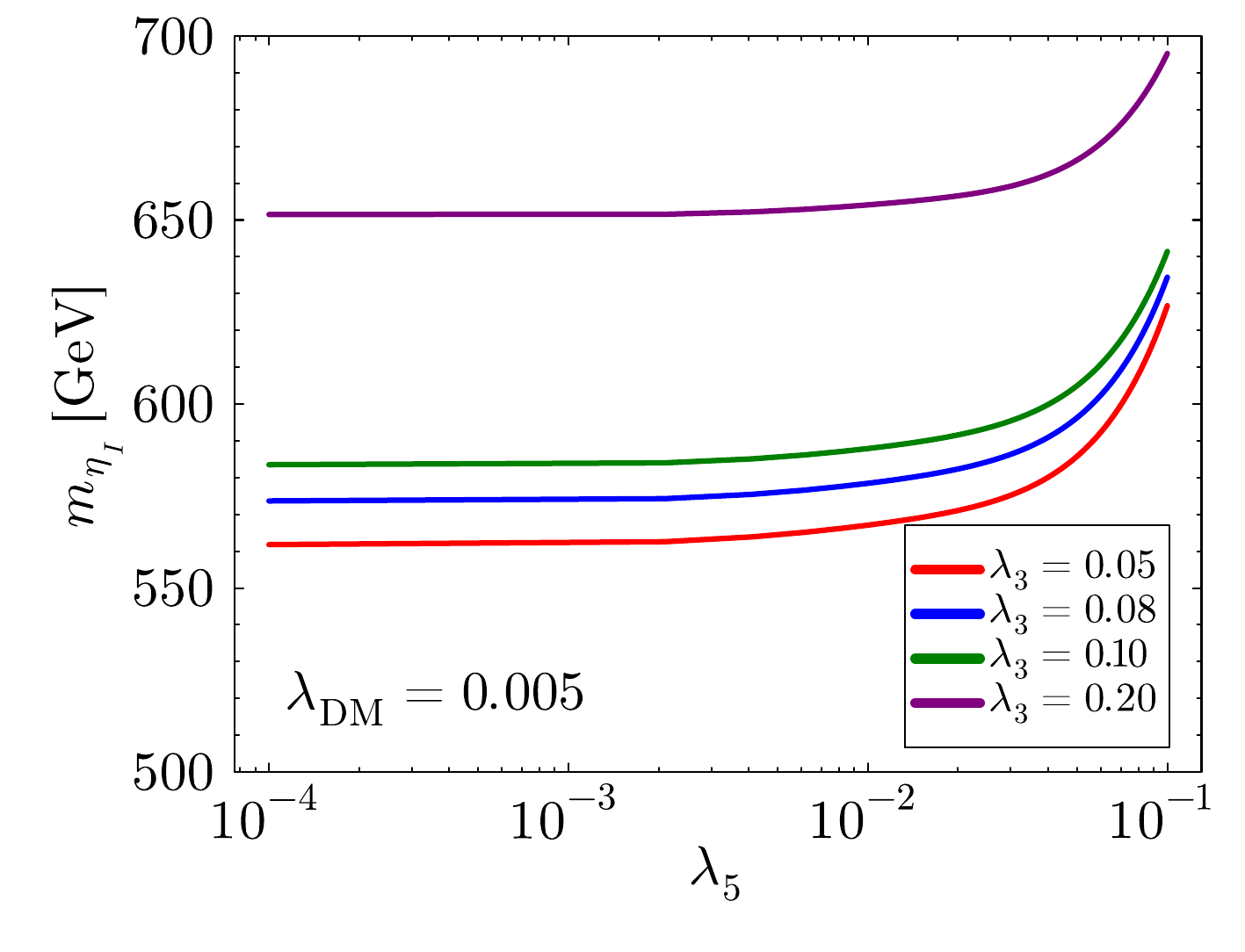}
	\includegraphics[width=0.45\textwidth]{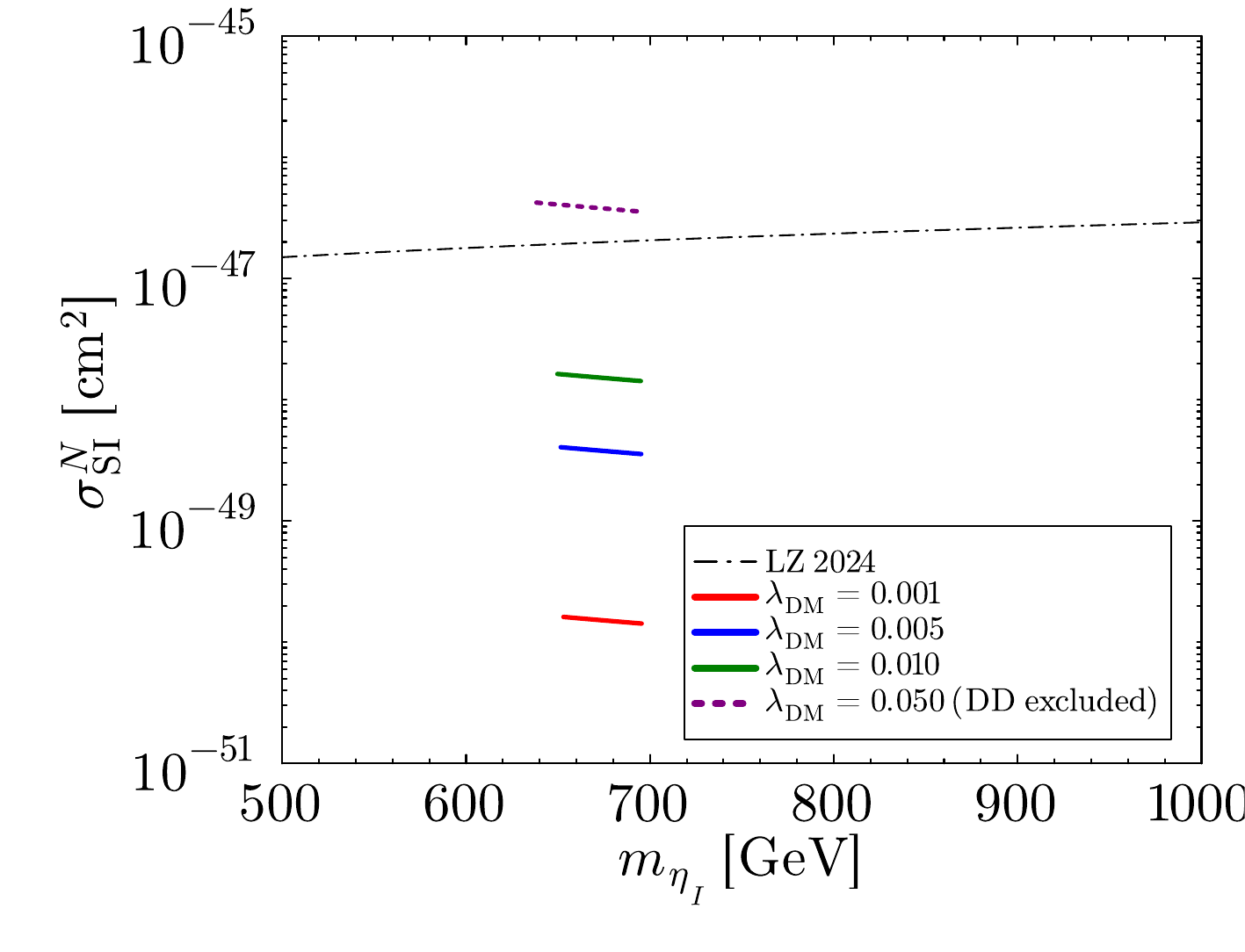}
	\caption{ Left panel: Variation of $m_{\ei}$ and $\lambda_5$. Here solid curves show the $\Omega_{ \ei} h^2 =0.12$ contours for different values of $\l_3$ when $\l_{\rm DM}=0.005$.  Right Panel: Variation of spin independent  DD cross-section of $\ei$ and $m_{\ei}$. Here, solid contours represent the $\Omega_{ \ei} h^2 =0.12$ contours for different values of $\l_{\rm DM}$ when $\l_3=0.2$. Dot-dashed contour represents the LUXZAPLIN 2024 bound on   DD cross-section. Purple dotted contour (for $\l_{\rm DM}=0.05$) is excluded for DD bound.}
	\label{fig:dm}
\end{figure}

The left panel of Fig.~\ref{fig:dm} shows the relic-density contours which  satisfies the $100\%$ DM relic density $\Omega_{\ei}h^2=0.12$ in the $(m_{\ei},\l_5)$ plane for fixed $\l_{\rm DM}=0.005$ and representative values of $\l_3$. 
As can be seen from the figure, for a fixed value of $\l_3$,  the value of $m_{\ei}$ required to reproduce the observed relic density increases with $\l_5$.  Generally, for a fixed value of $\l_3$, increasing $\l_5$ increases the neutral scalar splitting through $m_{\er}^2-m_{\ei}^2=\l_5v_\Phi^2$. This tends to reduce the thermal abundance of $\er$ in the coannihilating plasma. However, in the mass range shown the splitting remains only of order a few GeV, whereas the freeze-out temperature is typically $T_f\simeq m_{\ei}/25$. Thus $\er$ is not strongly Boltzmann suppressed and still participates efficiently in freeze-out. At the same time, increasing $\l_5$ enhances the scalar interactions involving $\er$, for example, the Higgs coupling $h\er\er$ contains the combination $\l_{\rm DM}+2\l_5$. In the region displayed in Fig.~\ref{fig:dm}, this coupling effect dominates over the modest Boltzmann suppression. As a result, the effective annihilation rate increases at fixed $m_{\ei}$. To restore $\Omega_{\ei}h^2=0.12$, one must then move to a larger $m_{\ei}$, where the heavy-WIMP annihilation rate is reduced approximately by the larger inert mass scale.

The same panel also shows that, for a fixed $\l_5$, increasing $\l_3$ shifts the relic-density contour toward larger $m_{\ei}$. This behavior has the same physical origin that at fixed $m_{\ei}, $ a larger $\l_3$ gives a smaller relic abundance. Therefore, a larger DM mass is needed to compensate it. 
Although $\l_3$ increases the charged-scalar mass through $m_{\eta_\pm}^2-m_{\ei}^2=(\l_3-\l_{\rm DM})v_\Phi^2/2$, the resulting reduction of charged coannihilation is not the full effect. The same parameter also changes scalar couplings, charged-scalar propagators, and the interference among the contact, $t/u$-channel, and Higgs-mediated diagrams contributing to channels such as $\ei\ei\to W^+W^-$ and $\ei\ei\to ZZ$. In fact, larger $\l_3$ increases the weight of pure gauge and gauge-scalar final states and can open or enhance channels such as $\eta^+\eta^-\to hh$, $\ei\eta^+\to W^+h$, and $\eta^+\er\to W^+h$. Hence the total effective annihilation rate increases with $\l_3$ in the displayed region, and the correct relic density is recovered only for larger $m_{\ei}$.

It is important to note that, in this heavy DM mass range, the spin-independent elastic scattering of $\ei$ off nucleons is controlled primarily by the Higgs-portal coupling $\lambda_{\rm DM}$. The corresponding Higgs-mediated DM-nucleon cross-section is given by~\cite{Barbieri:2006dq}
\begin{align}
	\sigma_{\rm SI}^{\mathcal{N}}= \frac{\lambda_{\rm DM}^2 f_{\mathcal{N}}^2}{4\pi}
	\frac{\mu_{\mathcal{N}}^2 m_{\mathcal{N}}^2}{m_h^4 m_{\ei}^2},
\end{align}
where $f_{\mathcal{N}}=0.32$ is the Higgs-nucleon coupling, $\mu_{\mathcal{N}}=m_{\mathcal{N}}m_{\ei}/(m_{\mathcal{N}}+m_{\ei})$ is the DM-nucleon reduced mass, $m_h\simeq125$ GeV is the SM Higgs mass, and $m_{\mathcal{N}}$ is the nucleon mass. Therefore, the non-observation of DM in DD experiments such as LUX, PandaX-II, XENON1T, and LUX-ZEPLIN mainly constrains $\lambda_{\rm DM}$ for a fixed DM mass, out of which LUX-ZEPLIN (LZ-2024)~\cite{LZ:2024vth} put the strongest constraint  as illustrated by the excluded part of the right panel of Fig.~\ref{fig:dm}.

The coupling $\lambda_5$ nevertheless remains crucial for direct searches through the inelastic $Z$-mediated process. Since $\lambda_5$ fixes the mass splitting between the CP-even and CP-odd neutral inert scalars, reducing $\lambda_5$ makes the transition $\ei+\mathcal{N}\to\er+\mathcal{N}$ kinematically easier. For a sufficiently small splitting, the halo DM kinetic energy can upscatter $\ei$ into the heavier state $\er$. For sub-TeV DM, splittings below roughly
$\Delta M\equiv m_{\er}-m_{\ei}\lesssim100$ keV would lead to a large inelastic scattering rate and are excluded by DD data~\cite{Arina:2009um,LopezHonorez:2010tb}. This implies the approximate lower bound $\Delta M\gtrsim100$ keV, or equivalently $\lambda_5\gtrsim 2m_{\ei}\Delta M/v_\Phi^2$, which is of order $10^{-6}$ for sub-TeV inert scalar DM.

To summarize, in this setup the correct DM relic abundance (assuming $\ei$ constitutes $100\%$ of the observed DM abundance) can be obtained for
\begin{align}
	550\lesssim \left( \frac{m_{\ei}}{\text{GeV}}  \right)\lesssim 1000,
	\qquad
	10^{-6}\lesssim\l_5\lesssim0.2 .
	\label{eq:bound}
\end{align} 
This region can be realized by varying $\l_3$ and $\l_{\rm DM}$ within the perturbative range, while keeping the Higgs-portal coupling sufficiently small to remain away from the DD exclusion bounds and satisfying the relevant indirect detection constraints.

Notice  that the above stringent bounds on $m_{\ei}$ and $\l_5$ can be relaxed if contribution of $\ei$ to DM relic is subdominant. In that case, the lower bound of $m_{\ei}$ and $\l_5$ in Eq.~\eqref{eq:bound} decreases and one can have
an inert scalar lighter than about $550\,{\rm GeV}$ for a fixed value of the quartic couplings, for instance, as can be shown in Fig.~\ref{fig:dmrelic}. As we will discuss in the proceeding sections, this can help us reduce the lightest RHN masses below TeV scale for the successful generation of baryon asymmetry.

 \section{Thermal leptogenesis in the minimal scotogenic model}
 \label{sec:lepto1}
 
 In this setup, RHN gets its mass below $B-L$ symmetry breaking. Its subsequent decay to (anti-)leptons and IHD can generate a finite amount of baryon asymmetry via thermal leptogenesis mechanism. In this case, a nonzero CP asymmetry is induced from the interference between the tree level as well as the one-loop level (including vertex and self-energy corrections) decay of the RHN, which is given by~\cite{Hugle:2018qbw}
 \begin{align}
 	\epsilon_i= \frac{1}{8\pi (Y^{\nu\dagger} Y^\nu)_{ii}} \sum_{j \neq i}  \text{Im}\left[ (Y^{\nu\dagger} Y^\nu)_{ij}^2\right]
 	F(x_{ij}, y_i),
 	\label{eq:cp1}
 \end{align}
 where $F(x_{ij},y_i)= \sqrt{x_{ij}}\left[1- \frac{1}{x_{ij}-1}(1-y_i)^2+ \frac{1-2y_i +x_{ij}}{(1-y_i)^2}\log\left(\frac{x_{ij}-y_i^2}{1-2 y_i + x_{ij}}\right)\right]$ is the loop factor with $x_{ij}\equiv M_j^2/M_i^2$ and $y_i\equiv m_\eta^2/M_i^2$.
 Further simplification of the CP asymmetry in Eq.~\eqref{eq:cp1} for a minimal two RHN system (using the CI parameterization) can be done in the limit $M_i\gg m_\eta$, which leads to the similar type-I seesaw structure~\cite{Davidson:2002qv} (although multiplied by the additional factor $\xi_2^{-1}$ indicating the scotogenic contribution)
 \begin{align}
 	\epsilon_1\simeq -\frac{3}{8\pi} \xi_2^{-1}  \frac{M_1}{v^2}  \frac{\sum_{\alpha} m_\alpha^2 \text{Im}\left[(R^*_{\alpha 1})^2\right]}{\sum_{\alpha} m_\alpha |R_{\alpha 1}|^2}.
 \end{align}
 Then using orthogonality condition $\sum_\alpha R^*_{\alpha 1}=1$,  one can evaluate the maximum CP asymmetry for Normal ordered light neutrino masses satisfying the condition~\eqref{eq:lightmass} as
 \begin{align}
 	|\epsilon_1| \lesssim \frac{3}{8\pi} \xi_2^{-1}  \frac{M_1}{v_\Phi^2} m_3,
 \end{align}
 where $m_3 \equiv \sqrt{\Delta m_{31}^2}$ is the largest light neutrino mass.
 
 With the knowledge of CP asymmetry, the baryon asymmetry can be evaluated from $Y_B=  \frac{C}{7.04}|\epsilon_1| \kappa_1$ where $C$ is the sphaleron conversion factor and $\kappa_1$ denotes the efficiency factor, which indicates the effect of washout of asymmetry~\cite{Davidson:2008bu}. The factor $\kappa_1$ depends on the decay parameter $K_1\equiv \Gamma_1/\mathcal{H}(T=M_1)$ where 
 \begin{align}
 	\Gamma_{i}= \frac{M_i}{8\pi} (Y^{\nu\dagger}Y^\nu)_{ii} (1-y_i)^2 ,
 \end{align}
 is the decay rate of the RHN to lepton and IHD above EWSB and  $\mathcal{H}= 0.33 \sqrt{g_*(T)} T^2/M_{p}$ 
 is the Hubble expansion rate of the radiation dominated Universe with $\mathcal{H}(T=M_1)$ denoting the expansion rate at $T=M_1$, $g_*$ is the eﬀective number of relativistic degrees of freedom and $M_{p} = 2.44\times 10^{18}$ GeV  is the reduced Planck mass.  The value of $K_1$ dictates whether the washout is strong (for $K_1>4$) or weak (for $K_1<1$) and analytically can be expressed for our setup ( in the limit $M_1>> m_{\eta}$) as
 \begin{align}
 	K_1= \frac{1}{4\pi v_{\Phi}^2}\xi_1^{-1}\sum_\alpha m_\alpha | R_{\alpha 1}|^2
 	\frac{M_{p}}{0.33 \sqrt{g_*(M_1)}}.
 	\label{eq:k1}
 \end{align}
 
 We note that since the decay parameter goes by $K_1\propto 1/\xi_1$, it depends on $\l_5$, which is correlated with the neutral inert scalar mass splitting entering the loop function for neutrino masses. More specifically, in the small mass splitting limit for neutral inert scalars,  $\xi_1\propto m_{\er}^2-m_{\ei}^2 \propto \l_5$ for a fixed value of $M_1$. Consequently,  one needs to fix the largest value of $\l_5$ allowed to get the minimum value of $K_1$. 
 
 Setting $\l_5=0.2$ (maximum allowed value from dark matter analysis) in Eq.~\eqref{eq:k1}, we observe that $K_1^{\rm min}$ is larger that $10^4$ for RHN mass $M_1=1$ TeV, while smaller value $\l_5$ would increase  $K_1^{\rm min}$. This leads to an important conclusion that the system will always be in the strong washout regime. In that case, the efficiency factor can be approximated as $\kappa_1\simeq 1/(1.2 K_1 [\log K_1]^{0.8})$~\cite{Buchmuller:2004nz}, which leads to the minimum value of the allowed lightest RHN mass  to be $M_1^{\rm min}\simeq 10^{11}$ GeV that generates the observed $Y_B=8.7 \times 10^{-11}$~\cite{Planck:2018vyg}.

 \section{Spontaneous leptogenesis from the Majoron}
 \label{sec:lepto2}
 
 The discussion in the previous section shows that  the conventional thermal leptogenesis in the minimal (two RHN) scotogenic model is pushed to a high scale. The reason is twofold. First, for hierarchical RHNs the CP asymmetry generated in $N_1$ decay is too small at low $M_1$. Second, the neutrino mass and DM motivated region is characterized by a very large decay parameter $K_1\gg 4$, so the system lies deep in the strong washout  regime. This strong washout erases most of the generated lepton asymmetry, thereby suppressing the efficiency of conventional thermal leptogenesis. Interestingly, it is the precise regime in which spontaneous leptogenesis becomes effective as fast $B-L$ violating interactions can track the equilibrium asymmetry induced by a time-dependent background. Let us elaborate in the following discussion.
 
 In the dynamical scotogenic model, spontaneous breaking of the global $U(1)_{B-L}$ symmetry produces the Majoron field $\theta$. After symmetry breaking, the RHN Majorana mass term takes the form
 \begin{align}
 	-\mathcal{L} \supset \frac{1}{2\sqrt{2}} Y^N_i v_\vp e^{i\theta} \overline{N_i^c} N_i +\text{h.c.}\,.
 \end{align}
 Then, the redefinition of the RHNs removes the phase in the Majorana mass terms and generates the derivative interactions between $\theta$ and $B-L$ current. Including the contribution induced by the RHN kinetic term, the relevant interactions therefore become \cite{Chun:2023eqc}
 \begin{align}
 	\mathcal{L} \supset -\frac12\partial_{\mu}\theta J^{\mu}_{B-L}-\frac14 \partial_{\mu}\theta \left(\overline{N^c_i}\gamma^{\mu}N_i^c-\overline{N_i}\gamma^{\mu}N_i\right).
 	\label{eq:deriv}
 \end{align}
 Notice that for a homogeneous background $\partial_\mu\theta\equiv (\dot{\theta},0,0,0)$, the second term in Eq.~\eqref{eq:deriv} gives rise to a helicity-dependent energy shift for the RHNs. However, in the strong washout regime, the asymmetry sourced by this helicity effect becomes strongly suppressed and can safely be neglected~\cite{Chun:2025abp}.
 We therefore focus on the universal derivative coupling to the $B-L$ current,
 \begin{align}
 	\mathcal{L}\supset -\frac12\dot{\theta} n_{B-L}, \qquad n_{B-L}= J^0_{B-L}.
 \end{align} 
 This interaction acts as an effective chemical potential for $B-L$ that shifts the equilibrium distributions of SM fermions and antifermions in opposite directions, so that the plasma develops a nonzero $B-L$ density whenever a $B-L$ violating interaction remains in thermal equilibrium.
 
 In the present setup, the required $B-L$ violation is mostly provided by the decay and inverse decay processes $N_1\leftrightarrow \ell_{L_\alpha} \eta$  below the $U(1)_{B-L}$ breaking scale. Since the same parameter region already satisfies $K_1\gg 4$, these reactions remain efficient for the extended period till $T\simeq M_1/10$. During this interval, the Majoron velocity (if nonzero) continuously biases the plasma towards a nonzero $B-L$ asymmetry. Once the inverse decay rate drops below the Hubble expansion rate, the relevant wash-in process freezes out, leaving the generated $B-L$ asymmetry conserved, up to the subsequent redistribution by spectator processes and electroweak sphalerons.

Depending on the neutrino Yukawa coupling, the scattering processes $\ell_{L_\alpha} \eta \leftrightarrow \bar{\ell}_{L_\alpha} \eta^\dagger$ can also source $B-L$ violation. The evolution of the lepton asymmetry, $n_{l_\alpha}\equiv (n_{\ell_{L_\alpha}}-n_{\bar{\ell}_{L_\alpha}})$ in this case can be described by the following Boltzmann equation in presence of the source $\dot{\theta}\neq 0$,  \cite{Domcke:2020kcp, Chun:2023eqc}
 \begin{align}
\Dot{n}_{\Delta l_{\alpha}}+3\mathcal{H}n_{\Delta l_\alpha}\simeq- \gamma_{\slashed{L}}	\left(\frac{2\mu_{l_\alpha}+2\mu_{\eta}}{T}-\frac{\Dot{\theta}}{T}\right)+...\,,	\label{eq:be}
 \end{align}
 where $\mu_{l_\alpha}$ and $\mu_\eta$ are the chemical potential for each component of lepton and $\eta$ doublet respectively.
 Moreover, $\gamma_{\slashed{L}}= \gamma_{N_i\to l_{L_\alpha}+\eta} + \gamma_{\Delta L=2}$ denotes the sum of thermally averaged interaction rate density for $N_1\to \ell_{L_\alpha} \eta$ decay and offshell $N_1$ mediated $\Delta L=2$  scattering $\ell_{L_\alpha}  \eta \to \bar{\ell}_{L_\alpha}  \eta^\dagger$. Here,
 \begin{align}
 	 \gamma_{ N_i\to l_{L_\alpha}+\eta}=
 	n_{N_1}^{\rm eq} \frac{\mathcal{K}_1(z)}{\mathcal{K}_2(z)} \Gamma_1, \qquad z=M_1/T,
 \end{align}
 with $\mathcal{K}_1,~\mathcal{K}_2$ are the Modified Bessel functions of first and second kind, respectively.
 
 The thermally averaged reaction density for the scattering process $\ell_{L_\alpha}\eta\rightarrow\bar{\ell}_{L_\alpha}\eta^*$, mediated by the $N_1$, is given by~\cite{Plumacher:1996kc}
 \begin{equation}
 	\gamma_{\Delta L=2}^{\rm off}(T)=
 	\frac{T}{64\pi^4}
 	\int_{s_{\rm min}}^\infty ds
 	\sqrt{s}\,
 	\mathcal{K}_1\left(\frac{\sqrt{s}}{T}\right)
 	\hat{\sigma}_{\Delta L=2}^{\rm off}(s),
 	\label{eq:gamma_DL2_off}
 \end{equation}
  $s_{\rm min}=m_\eta^2$ in the approximation of massless leptons, and $\hat{\sigma}_{\Delta L=2}^{\rm off}(s)$ represents the reduced cross section after subtracting the real intermediate state (RIS) contribution associated with on-shell $N_1$ exchange.
 
 Remember that the RIS subtraction is necessary to avoid double counting the contribution already included through the decay and inverse-decay processes $N_i\leftrightarrow\ell_{L_\alpha}\eta$. Schematically, the RIS-subtracted reduced cross section is defined as
 \begin{equation}
 	\hat{\sigma}_{\Delta L=2}^{\rm off}(s) 	=
 	\hat{\sigma}_{\Delta L=2}^{\rm full}(s)
 	-
 	\hat{\sigma}_{\Delta L=2}^{\rm RIS}(s).
 	\label{eq:sigma_DL2_off}
 \end{equation}
 Here, the RIS contribution corresponds to the resonant part of the $s$-channel amplitude, evaluated consistently within the narrow-width approximation. The nonresonant contributions, including the $t$-channel exchange and the off-shell part of the $s$-channel exchange, are retained in $\hat{\sigma}_{\Delta L=2}^{\rm off}(s)$~\cite{Takahashi:2026ngu}.

 As long as the either  $B-L$ violating process remains in thermal equilibrium, the right hand side of Eq.~\eqref{eq:be} vanishes, implying
 \begin{align}
 	\mu_{\ell_\alpha}+\mu_{\eta} = \frac{\dot{\theta}}{2},  
 \end{align}
 Together with the chemical equilibrium conditions imposed by the other SM interactions, this relation
 then induces a nonzero baryon asymmetry
 \begin{align}
 	Y_B= \frac{\mu_B}{6} \frac{T^2}{s}= \frac{c_B}{6}Y_\theta \left(\frac{M_1}{v_\vp z}\right)^2, \qquad Y_\theta\equiv \frac{v_\vp^2 \dot{\theta}}{s},
 	\label{eq:yb}
 \end{align}
 where $\mu_B= c_{\rm sp} \mu_{B-L} $ with $c_{\rm sp}=8/23$ being the sphaleron conversion factor connecting nonzero $B-L$ and $B$ asymmetry and $\mu_{B-L}=c_{B-L} \dot{\theta}$. The coefficient $c_{B-L}$ and the effective baryon conversion factor $c_B=c_{\rm sp} c_{B-L}$ are determined by the chemical equilibrium conditions in the scotogenic plasma, as summarized in the Appendix. The baryon asymmetry in Eq.~\eqref{eq:yb} saturates at the decoupling temperature $T_{\rm dec}\equiv {\rm min}(T_{\rm dec}^{\rm inv},T_{\rm dec}^{\rm sc})$, where $T_{\rm dec}^{\rm inv}$ is determined from the condition
 \begin{align}
 	\langle \Gamma_{ID}\rangle/\mathcal{H}=1,~\text{with}~
 	\langle\Gamma_{ID}\rangle= \frac{n_{N_1}^{\rm eq}}{n_{\ell}^{eq}} \langle\Gamma_{N_1\to \ell_{L_\alpha} \eta}\rangle,
 \end{align}
 whereas  $T_{\rm dec}^{\rm sc}$ is evaluated from the condition $\gamma_{\Delta L=2}^{\rm off}/(n_\ell^{\rm eq} \mathcal{H})=1$.
 The final baryon asymmetry is then denoted as
 \begin{align}
 	Y_B= \frac{c_B}{6}Y_\theta \left(\frac{M_1}{v_\vp z_{\rm dec}}\right)^2,
 	\label{eq:yb1}
 \end{align}
 where $z_{\rm dec}=M_1/T_{\rm dec}$.  
 
 The $B-L$ violating processes can decouple before or after the electroweak sphaleron decoupling temperature $T_{\rm sp}$, which leads to two interesting possibilities. 
 If $T_{\rm dec}> T_{\rm sp}\simeq 132$ GeV, the final $Y_B$ saturates to the expression~\eqref{eq:yb1}. On the other hand, if $B-L$ violation remains active below $T_{\rm sp}$ (for TeV scale RHNs), the baryon asymmetry is instead fixed at $T_{\rm sp}$
 \begin{align}
 	Y_B= \frac{c_B}{6}Y_\theta \left(\frac{M_1}{v_\vp z_{\rm sp}}\right)^2,\qquad z_{\rm sp}= M_1/T_{\rm sp}.
 	\label{eq:yb2}
 \end{align}
 Here, the initial Majoron motion, characterized by a nonvanishing $Y_\theta$, may originate from an explicit breaking of the global $U(1)_{B-L}$ symmetry in the early Universe when $\varphi\gg v_\vp$. 
 Later, when $\varphi$ starts approaching $ v_\vp$,  explicit symmetry-breaking effects become negligible.  The equation of motion for the angular field then implies
 $\frac{d}{dt}\left(a^3\varphi^2\dot{\theta}\right)\simeq 0,$
 where $a$ denotes the cosmological scale factor. Consequently, assuming entropy conservation, the Majoron charge yield
 $Y_\theta\equiv\frac{n_\theta}{s}
 =\frac{\varphi^2\dot{\theta}}{s}
 $		
 remains approximately constant. 
 In the following, however, we remain agnostic about the underlying dynamics responsible for generating the initial condition for the Majoron motion, but assume that the additional dynamics do not influence the late-time evolution of the Majoron before $T_{\rm dec}$ \cite{Co:2019wyp,Lee:2023dtw,Lee:2024bij,Chun:2025brc}. Thus, we treat $Y_\theta$ as a free parameter in our work. 
 
We also note that the interactions with the thermal bath could damp the Majoron rolling during the evolution down to $T_{\rm dec}$. In the present setup, however, since the inverse decay process and $N_1$ mediated scattering process become active around $M_1\sim T \ll v_\vp$, the dissipation of the Majoron motion remains suppressed as $T^2/v_{\varphi}^2$~\cite{Domcke:2020kcp, Co:2019wyp, Domcke_2022} and one can treat $Y_\theta$ as an approximately conserved quantity. However, the situation can be different for processes involving the heavier RHN $N_2$, whose mass is taken to be $M_2\sim v_\varphi$. If the thermal bath reaches temperatures comparable to $M_2$, these interactions may become efficient without a significant suppression of the Majoron charge transfer. Therefore, they could potentially affect the conservation of $Y_\theta$. To avoid this possibility, we will explicitly impose the additional condition that the maximum temperature attained by the thermal bath satisfies $T_{\rm max}\ll v_\varphi\sim M_2$. Under this assumption, the thermal production and inverse decays of $N_2$ are Boltzmann suppressed. Moreover, the off-shell $N_2$-mediated $\Delta L=2$ scattering processes are suppressed at temperatures well below $M_2$. Thus, provided that these off-shell contributions are negligible, the dissipation of $Y_\theta$ associated with $N_2$ can be safely neglected.

 An interesting feature of the scotogenic leptogenesis is that there could be a nonzero chemical potential for IHD, $\mu_\eta$, whose evolution is controlled by the $\l_5$ interaction
 \begin{align}
 	V\supset \frac12 \l_5 \left[(\eta^\dagger \Phi)^2 +~\text{h.c.}\right].
 \end{align}
 Sufficiently large value of $\l_5$ can equilibrate this interaction, which leads to the condition $\mu_H= \mu_\eta$. It then changes the spectator coefficients $c_{B-L}$ relative to the ordinary type-I seesaw case (see Appendix). 
 For example, with $\l_5\sim 10^{-2}$ the interaction effectively remains efficient over a wide temperature range, extending from very high scales down to temperatures below $T_{\rm sp}$. In that case, one needs to impose the condition $\mu_H= \mu_\eta$ while calculating $c_{B-L}$\footnote{ Since different SM Yukawa and spectator interactions enter chemical equilibrium at different temperatures, the set of equilibrium conditions relevant for redistributing the generated asymmetry depends on $T_{\rm dec}$. Consequently, the corresponding conversion coefficient $c_{B-L}$ also varies with $T_{\rm dec}$.}. For smaller values, such as $\l_5\sim 10^{-4}$, the interaction enters equilibrium only below $T_{\rm eq}^{\l_5}\sim 5 \times 10^6$ GeV and remains in equilibrium below $T_{\rm sp}$. For even smaller $\l_5$, which can be allowed if $\ei$ constitutes only a subdominant fraction of DM, the interaction may never reach thermal equilibrium. Accordingly, if the decoupling of $B-L$ violating process occurs before the $\l_5$ interaction becomes efficient, or if the latter never equilibrates, the chemical equilibrium condition relating the inert doublet to the Higgs sector cannot be imposed, and $\mu_\eta$ must instead be treated as an independent chemical potential.

 Finally,  we comment on a further cosmological aspect of the present setup. Since the Majoron may also acquire a small mass from the explicit breaking of the global $U(1)_{B-L}$ symmetry, its coherent motion can contribute to the present DM abundance. It is therefore necessary to ensure that this contribution does not overclose the Universe. 
 The estimation of the resulting Majoron relic abundance requires the knowledge of the initial energy stored in the coherent Majoron field.
 A simple and well-motivated possibility is that the global $U(1)_{B-L}$ symmetry is explicitly broken at the Planck scale, as expected from quantum-gravity effects~\cite{Kamionkowski:1992mf,Kallosh:1995hi}.  In the early Universe, when the radial field of $\vp$ is displaced from its $vev$,  this explicit symmetry breaking generates a potential along the angular direction and can source a nonzero initial velocity for the Majoron field \cite{Lee:2023dtw,Lee:2024bij,Chun:2025brc}. 
 As the Universe expands, the radial field subsequently relaxes toward its vacuum value $\vp= v_\vp/\sqrt{2}$. However, the Majoron keeps on rolling till it gets trapped in its potential. We assume the Majoron potential to have the form
 \begin{align}
 	V(J)= m_J^2 v_{\vp}^2 \left[1-\cos\left(\frac{J}{v_\vp}\right)\right],
 	\label{eq:cosine}
 \end{align}
 where $m_J$ is the Majoron mass.
 
 In this case, if the kinetic energy stored in the angular direction exceeds the potential barrier $V_{\cancel{B-L}}^{\rm max}=2m_J^2 v_\vp^2$ at $m_J\simeq 3 \mathcal{H}(T_{\rm osc})$, the onset of the usual Majoron oscillation is delayed as the Majoron keeps rolling until its kinetic energy redshifts down to the height of the potential barrier, i.e., $\rho_\theta(T_{\rm trap})= V_{\cancel{B-L}}^{\rm max}$, through the kinetic misalignment~\cite{Co:2019wyp, Chun:2025brc}. In terms of the conserved angular yield $Y_\theta$, the regime for kinetic misalignment is achieved for
 
 \begin{align}
 	Y_{\theta} \quad \gtrsim 4.51\, \frac{g_*({T_{\rm osc}})^{3/4}}{g_*^s({T_{\rm osc}})} \bigg(\frac{v_\vp^2}{m_J^{1/2}M_p^{3/2}}\bigg) \equiv Y_{\text{cr}}
 	\label{eq:ycr}
 \end{align}
 where 
 \begin{align}
 	T_{\text{osc}}=\left(\frac{m_J M_p}{0.99\sqrt{g_*}}\right)^{\frac12},
 \end{align}
 which is the temperature where conventional oscillation would begin.\footnote{For $Y_{\theta}<Y_{\text{cr}}$, the Majoron would have already been trapped before the onset of oscillation. In this regime (also known as the misalignment mechanism~\cite{Preskill:1982cy, Abbott:1982af,Dine:1982ah}), the Majoron relic abundance depends on the displacement from the potential minimum of $V_{\cancel{\text{B-L}}}$. A reliable determination of the resulting abundance requires knowledge of the initial condition for the spontaneous $B-L$ breaking, which can be made before or after inflation, so we don't pursue the details of the misalignment mechanism in our work. We thus identify only the kinetic misalignment regime for the Majoron dark matter and leptogenesis process, namely, $Y_{\theta}>Y_{\text{cr}}$~\cite{Co:2019jts, Chun:2023eqc}.}.
 
 The Majoron abundance produced by the angular motion is controlled by the same yield $Y_\theta$\footnote{A large Majoron velocity can source inhomogeneous Majoron perturbations which eventually can contribute as an extra source for Majoron relic density~\cite{Eroncel:2025qlk, Bodas:2025eca}. However, for low Majoron velocity regime, which can be achieved for lower values of $v_\varphi$, this effect will be subdominant.}. In that case, if the Majoron were to account for the observed DM abundance, the required yield would be \cite{Co:2019jts}
 \begin{align}
 	Y_{\theta}^{\rm DM}\approx \ \frac{0.44 \text{eV}}{2 m_J}.
 	\label{eq:kindm}
 \end{align}
 
 Assuming that $\ei$ from the IHD sector constitutes a negligible fraction of the observed DM abundance, the Majoron contribution is constrained not to exceed the remaining DM abundance. This translates into an upper bound on the Majoron yield,
 \begin{align}
 	Y_\theta \leq Y_\theta^{\rm DM}.
 	\label{eq:dmcond2}
 \end{align}
 Thus, the parameter region relevant for spontaneous leptogenesis should be chosen in such a way that the rolling Majoron generates the required effective chemical potential for $B-L$ while the Majoron relic abundance (along with the $\ei$ abundance) remains compatible with the observed DM abundance.

\section{Parameter space for scoto-leptogenesis}
\label{sec:lepto3}

\begin{figure}[h]
	\includegraphics[width=0.45\textwidth]{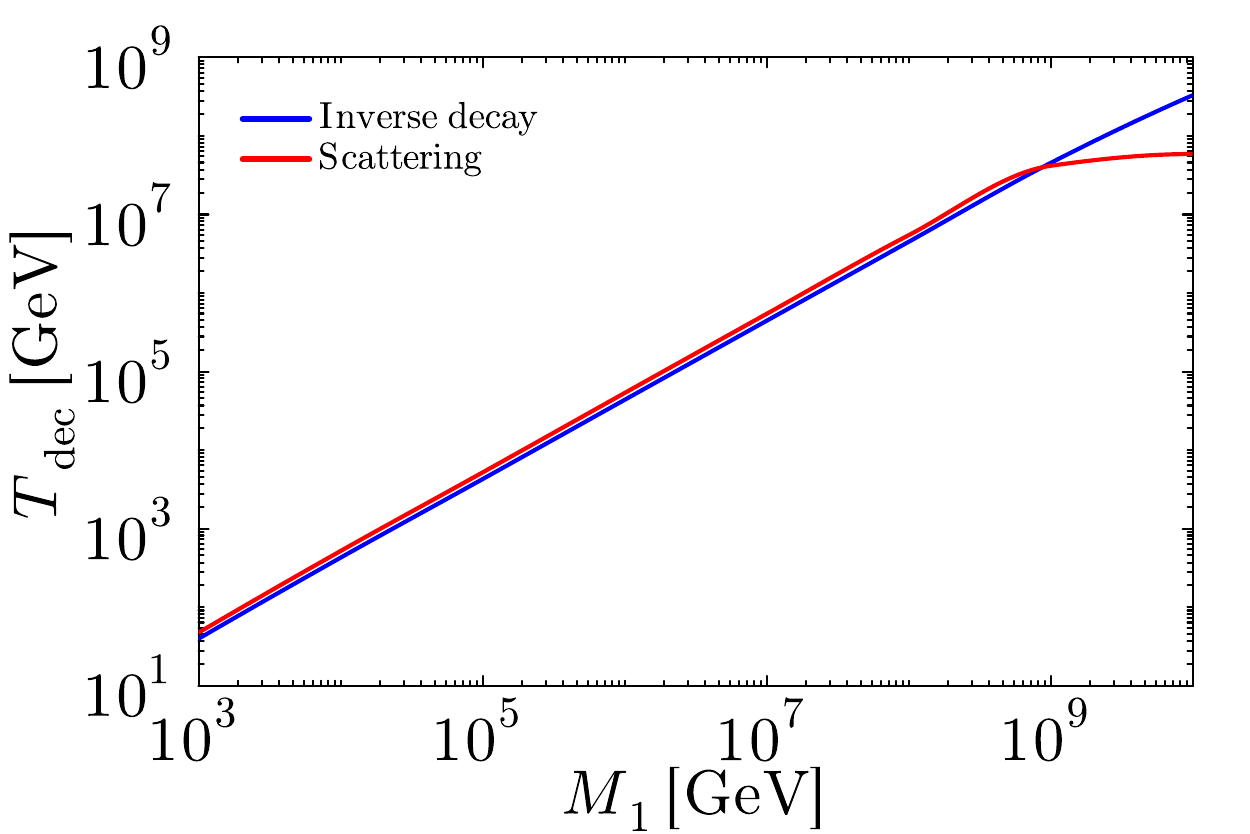}
	\includegraphics[width=0.45\textwidth]{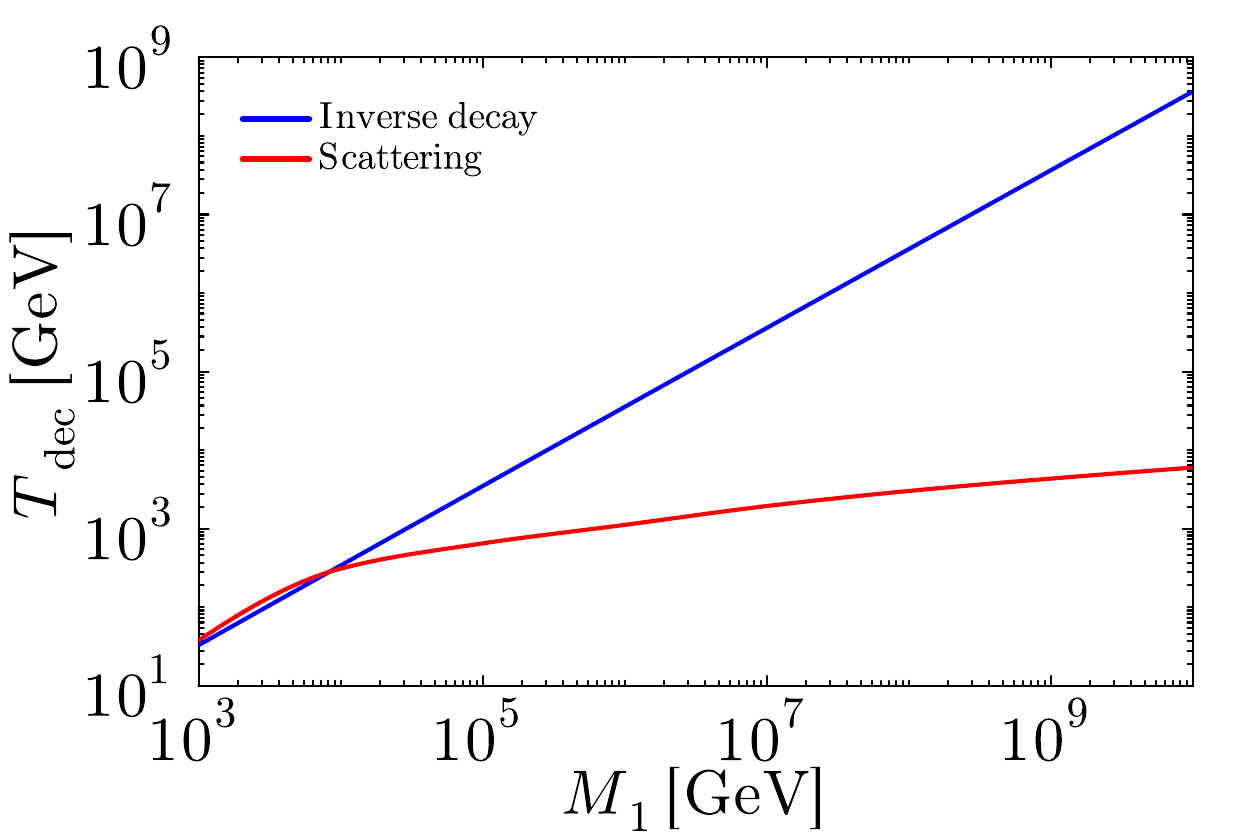}
	\caption{Variation of decoupling temperature for inverse decay and $\Delta L=2$ offshell $l_L\eta\to \bar{l}_L \eta^\dagger$ scattering processes w.r.t.  $M_1$ for $\l_5=10^{-2}$ (left panel) and $\l_5=10^{-4}$ (right panel) respectively.}
	\label{fig:dec}
\end{figure}

\begin{figure}[h]
	\includegraphics[width=0.45\textwidth]{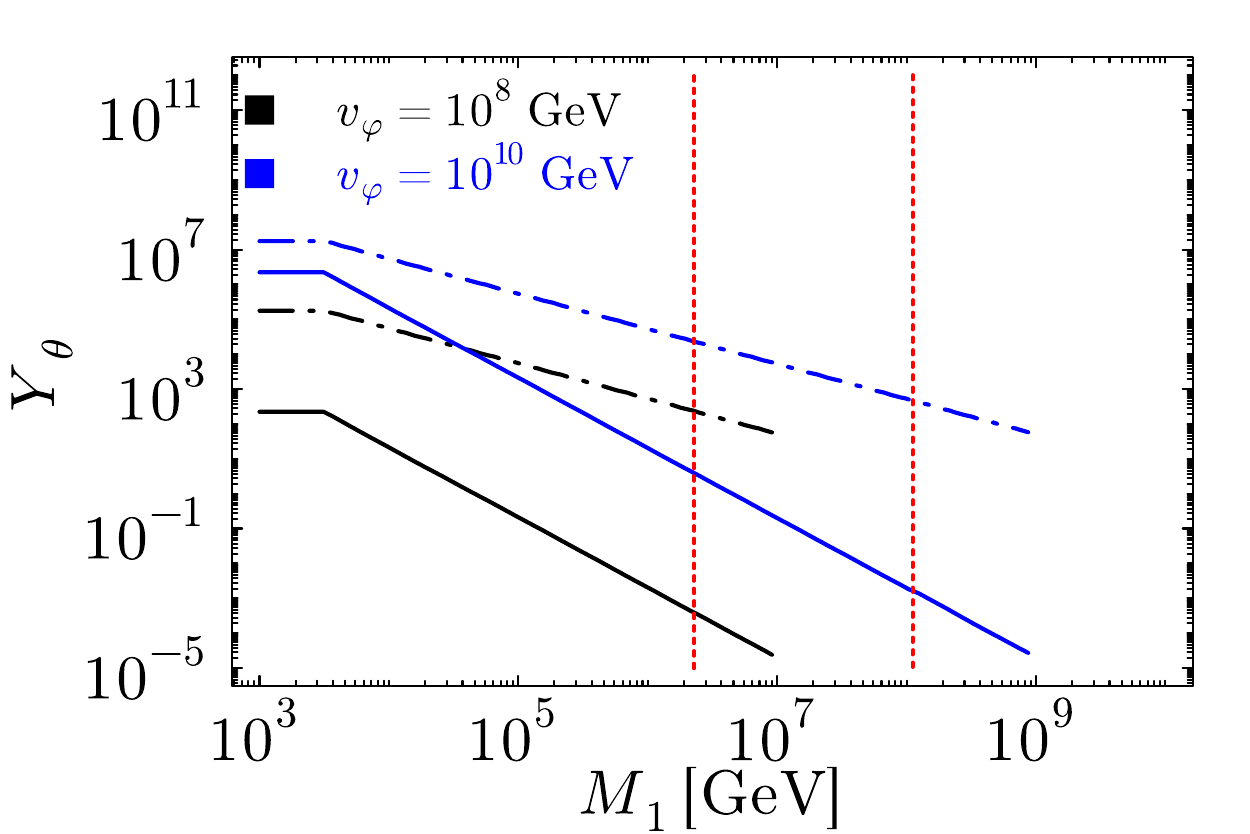}
	\includegraphics[width=0.45\textwidth]{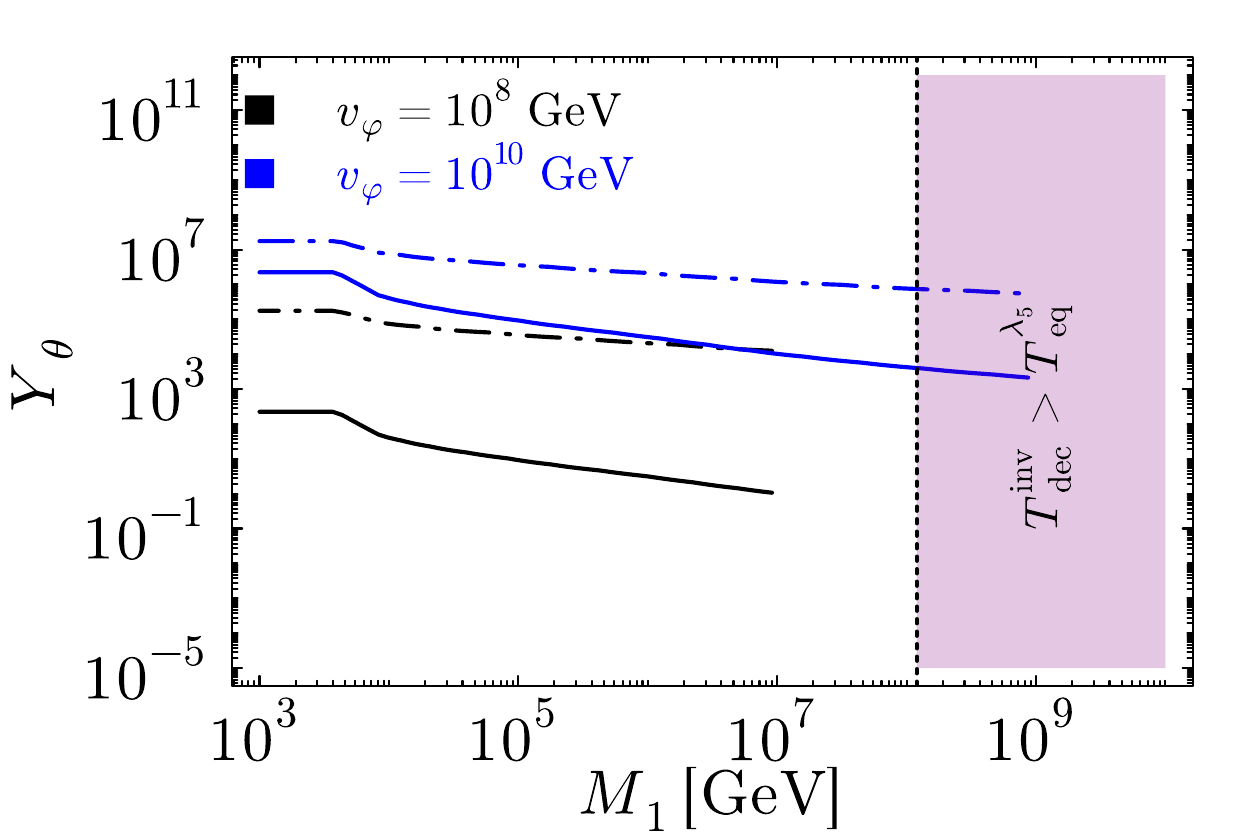}
	\caption{Parameter space for $Y_{\theta}$ vs $M_1$, with  $\lambda_5=10^{-2}$ (left panel) and $10^{-4}$ (right panel). We set the Casas-Ibarra angle as $\vartheta=\frac{\pi}{6}+i$. Black and blue solid lines represent available parameter set, $\{Y_{\theta},M_1\}$, that generates current baryon asymmetry for $v_{\varphi}=10^{8}$ GeV and $v_{\varphi}=10^{10}$ GeV respectively. The black and blue dot-dashed lines represent the value of $Y_\theta$ for which $\rho_{\theta,{\text{inv}}}^{\rm dec}= \rho_{R,{\text{inv}}}^{\rm dec}$ for  $v_{\varphi}=10^{8}$ GeV and $v_{\varphi}=10^{10}$ GeV, respectively. }
	\label{fig:bau1}
\end{figure}

\begin{figure}[t]
	\includegraphics[width=0.45\textwidth]{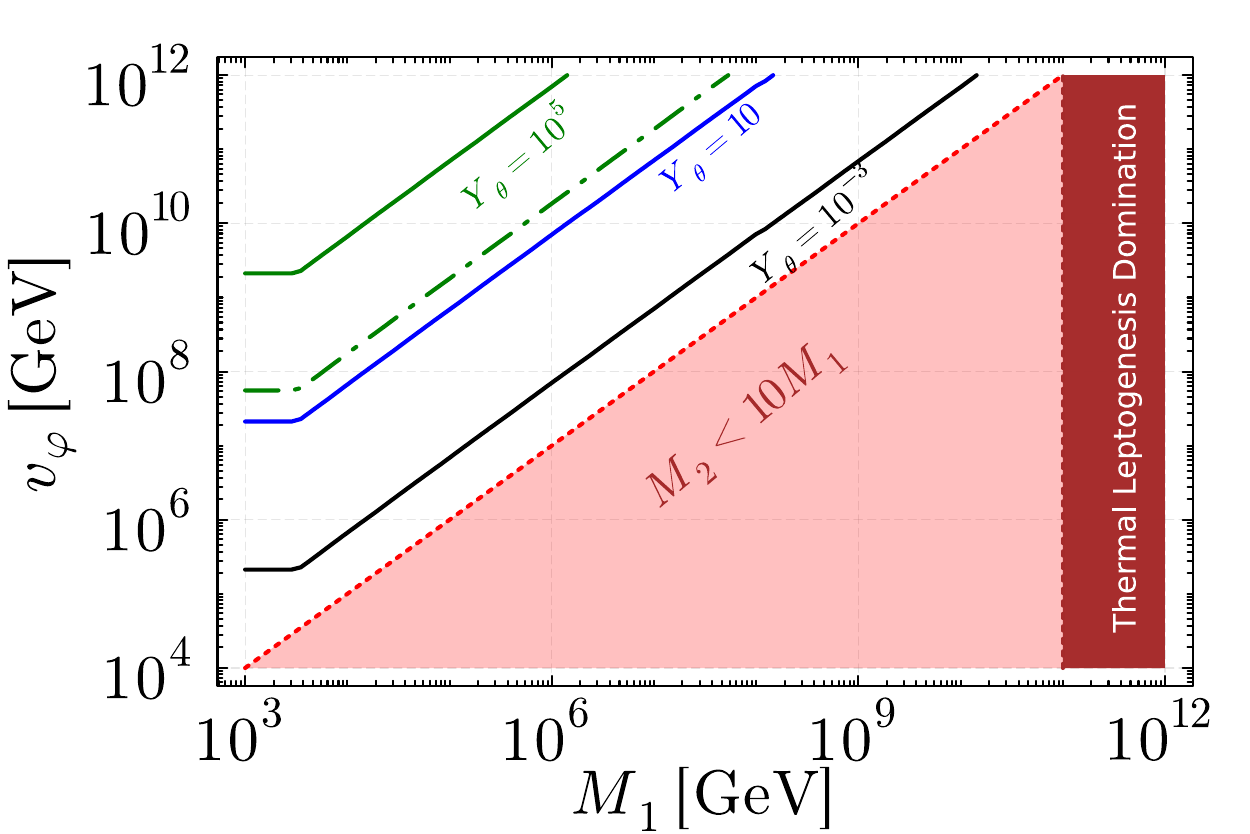}
	\includegraphics[width=0.45\textwidth]{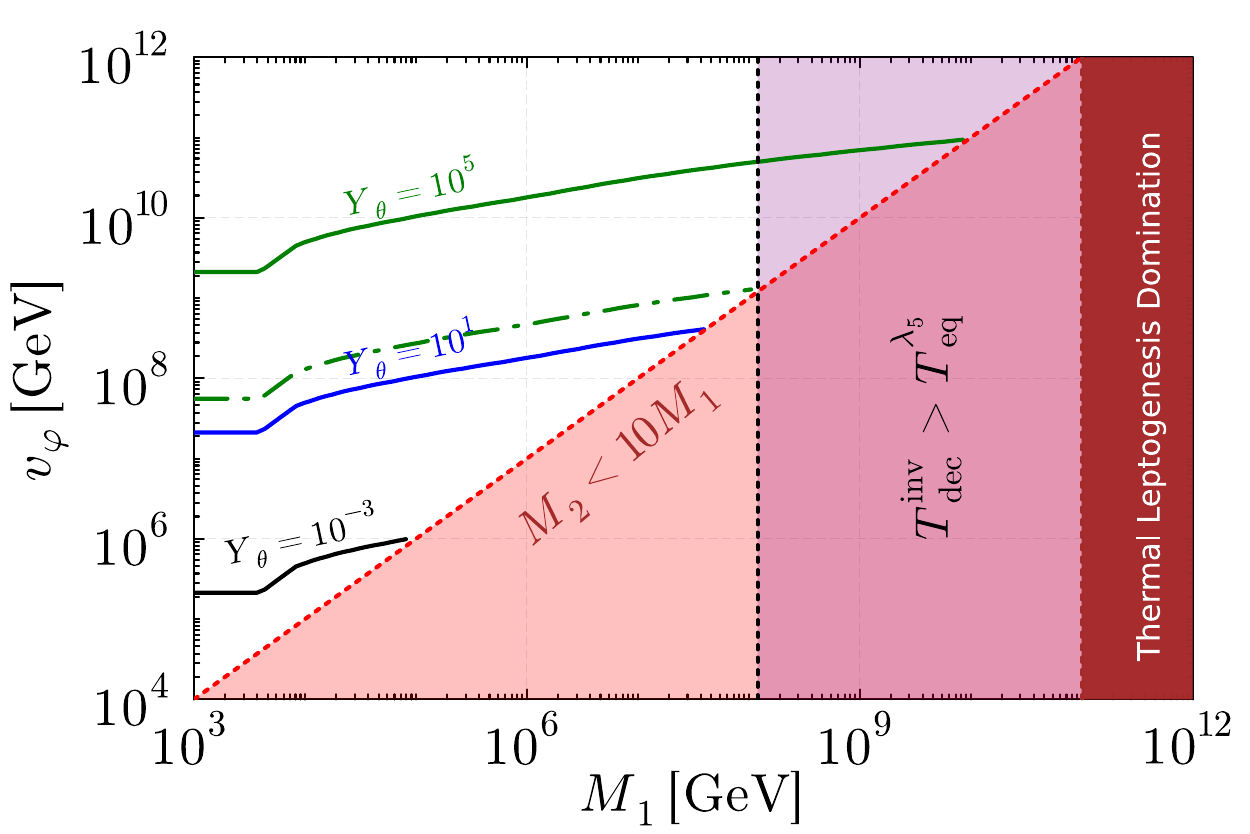}
	\caption{Parameter space for successful scoto-leptogenesis in $v_{\phi}$ vs $M_1$. We chose $\lambda_5=10^{-2}$, $\lambda_5=10^{-4}$, on left and right figures, respectively. We set the Casas-Ibarra angle as $\vartheta=\frac{\pi}{6}+i$. Solid lines indicate a set of correct parameters satisfying the observed $Y_B$ for $Y_{\theta}=10^{-3}, 10, 10^{5}$.}
	\label{fig:bau2}
\end{figure}

We now present the parameter region in which the rolling Majoron background produces the observed baryon asymmetry while remaining consistent with the neutrino mass and DM constraints previously discussed. Throughout this section, we take the CP-odd inert scalar $\ei$ and/or the Majoron to be the viable DM candidates and use the heavy IHD region identified in Sec.~\ref{sec:dm}. Unless otherwise stated, we also fix the Casas-Ibarra angle to $\vartheta= \pi/6+i$.  
Even in the range $\vartheta_i\lesssim 1$, the decay parameter $K_1$ remains deep in the strong washout regime for two hierarchical RHNs (see associated discussion in  Sec.~\ref{sec:lepto1}). Consequently, changing $\vartheta_i$ only mildly affects the inverse decay decoupling temperature, which remains around $z_{\rm dec}=\mathcal{O}(20)$ for the benchmark points considered here, as well as for a more general range of $\vartheta$. Finally, since the RHN mass is generated due to the nonzero $vev$ of $\vp$, exact estimates of $M_i$ require the knowledge of $Y^N_i$. For the following discussion, we will set $Y^N_2=1$ so that $M_2=v_{\vp}$ while $Y^N_1$ will be varied freely as long as the condition $M_2<10 M_1$ is not violated to maintain a conservative hierarchical structure of RHN masses.

Following these assumptions, in Fig.~\ref{fig:dec}, we show the decoupling temperatures of inverse decays of $N_1$ and the $\Delta L=2$ off-shell scattering process $\ell_L\eta\to\bar{\ell}_L\eta^\dagger$ mediated by  $N_1$ as functions of the lightest RHN mass $M_1$. The left and right panels correspond to two representative choices, $\lambda_5=10^{-2}$ and $10^{-4}$, respectively. As evident from the figure, for sufficiently large $M_1$, the off-shell scattering process remains in thermal equilibrium down to temperatures significantly below the inverse decay decoupling temperature. This behavior can be understood from the different dependence of the two processes on the neutrino Yukawa couplings and their temperature evolution. In particular, the inverse decay rate is proportional to $(Y^{\nu \dagger} Y^\nu)_{11}$, whereas the $N_1$-mediated $\Delta L=2$ scattering rate scales as $(Y^{\nu \dagger} Y^\nu)_{11}^2$, with additional dependence on the RHN mass and thermal kinematics. For fixed light-neutrino masses and CI parameters, $(Y^{\nu \dagger} Y^\nu)_{11}$ generally increases with $M_1$, enhancing the scattering amplitude. Moreover, unlike inverse decays, which become exponentially suppressed at $T\ll M_1$ due to the decreasing equilibrium abundance of the RHN, off-shell scattering can proceed through virtual RHN exchange without such suppression by the RHN abundance. Consequently, the scattering process can remain effective well after inverse decays have decoupled. This effect becomes more pronounced for smaller $\l_5$, since $(Y^{\nu \dagger} Y^\nu)_{11}$ scales approximately as $1/\l_5$ for fixed $M_1$ and CI parameters. The enhanced Yukawa couplings allow $\Delta L=2$ scattering to remain in equilibrium below the inverse decay decoupling temperature even for comparatively smaller $M_1$, as illustrated in the right panel of Fig.~\ref{fig:dec}.

The implications for baryogenesis are illustrated in Fig.~\ref{fig:bau1}, where we show the Majoron charge yield $Y_\theta$ required to reproduce the observed baryon asymmetry as a function of $M_1$ for the same two choices of $\l_5$. As long as the $B-L$-violating interactions remain efficient, the plasma tracks the equilibrium asymmetry induced by the effective chemical potential associated with $\dot{\theta}$. The final baryon asymmetry is determined when either the $B-L$-violating interactions or electroweak sphalerons decouple, whichever occurs first as the Universe cools. The corresponding temperature is therefore given by $T_d=\max(T_{\rm dec},T_{\rm sp})$, where $T_{\rm dec}=\min(T_{\rm dec}^{\rm inv},T_{\rm dec}^{\rm sc})$. Since the resulting baryon asymmetry scales as $Y_B\propto Y_\theta T_d^2/v_\varphi^2$, the different decoupling behaviors lead to distinct features in the two panels.

For $\l_5=10^{-2}$ (left panel), inverse decays determine the effective decoupling temperature over most of the relevant RHN mass range. In particular, for $M_1\gtrsim3\times10^3$ GeV, the $B-L$-violating interactions decouple above the sphaleron decoupling temperature. As $M_1$ decreases, the inverse decay decoupling temperature also decreases, requiring a larger $Y_\theta$ to reproduce the observed baryon asymmetry. Once $M_1\lesssim3\times10^3$ GeV, the $B-L$-violating interactions remain effective below $T_{\rm sp}$, and the final asymmetry is instead fixed at sphaleron decoupling. This gives rise to the plateau in the required $Y_\theta$, which becomes independent of $M_1$ for fixed $v_\varphi$.

For $\l_5=10^{-4}$ (right panel), the enhanced Yukawa couplings make the off-shell $\Delta L=2$ scattering process more important. In this case, scattering remains effective below the inverse decay decoupling temperature over a wider range of $M_1$, shifting the effective $B-L$ decoupling temperature to lower values. Consequently, a larger Majoron charge yield is required to reproduce the observed baryon asymmetry compared to the $\l_5=10^{-2}$ case, provided that the effective decoupling temperature remains above $T_{\rm sp}$. Once the scattering decoupling temperature falls below $T_{\rm sp}$, the final asymmetry is again determined by sphaleron decoupling, and the dependence on the scattering decoupling temperature disappears.

In both figures, the dot-dashed curves here indicate the boundary at which the kinetic energy density stored in the Majoron equals the radiation energy density at $T_d$. Above these curves, the Universe would enter a kination dominated phase before the baryon asymmetry freezes out. Since the analytic treatment in Sec.~\ref{sec:lepto2} assumes radiation domination, we restrict our numerical results to the regions below this boundary. Notice that this requirement becomes more restrictive for a larger $v_\vp$, because the same value of $Y_\theta$ generates a larger kinetic energy in the angular mode with a larger $v_\vp$. As a result, successful low-scale leptogenesis is most easily realized for moderately small symmetry breaking scales\footnote{The region above the dot-dashed curves requires a dedicated analysis including the effect of equilibration of interactions in the modified expansion history. For comparison, in thermal leptogenesis with kination domination \cite{Chun:2007np}, it was shown that the strong wash-out regime in radiation domination turns into a less strong wash-out regime, because the Hubble parameter becomes larger before decoupling. Nonetheless, our discussion in the strong wash-out regime would remain valid in most of the parameter space.}.

Finally, we discuss the role of the $\l_5$ interaction in determining the spectator effects and the corresponding conversion coefficient $c_B$. For $\l_5=10^{-2}$, this $\l_5$ interaction remains in thermal equilibrium throughout the temperature range relevant for inverse decay decoupling, enforcing the chemical-equilibrium condition $\mu_\eta=\mu_H$. As discussed in the previous section, this modifies the coefficient $c_B$ relative to the conventional type-I seesaw scenario. Interestingly, the same equilibrium condition also holds for the smaller coupling $\l_5=10^{-4}$, although for a different reason. In this case, the $\l_5$ interaction enters thermal equilibrium only below $T\simeq5\times10^6$ GeV. Therefore, for sufficiently large $M_1$, inverse decays may decouple before this interaction becomes efficient (indicated by the purple shaded region in the right plot of Fig.~\ref{fig:bau1}). However, as shown in Fig.~\ref{fig:dec}, the $\Delta L=2$ off-shell scattering process remains in equilibrium down to temperatures well below the inverse decay decoupling temperature. Consequently, by the time the $B-L$-violating interactions finally decouple, the $\l_5$ interaction has already reached thermal equilibrium and enforces $\mu_\eta=\mu_H$. Thus, despite the different equilibration temperatures of the $\l_5$ interaction, the same spectator coefficient $c_B$ applies to both choices of $\lambda_5$ in the parameter region considered.

In Fig.~\ref{fig:bau2}, we illustrate the parameter space for the successful leptogenesis in the $(M_1,v_\vp)$ plane for fixed values of the Majoron yield $Y_\theta$. The contours show that the observed $Y_B$ can be obtained over a wide range of RHN masses, including the TeV scale, without requiring the quasi-degenerate mass spectrum. Similar to Fig.~\ref{fig:bau1}, the right side of the dot-dashed contours indicates kination domination. So, a very large $Y_\theta$ is  incompatible with the radiation domination assumption. In both plots of Fig.~\ref{fig:bau2}, the red shaded regions indicate $M_2<10 M_1$, which violates the underlying assumption for hierarchical RHN masses. For the case of $\l_5=10^{-4}$,  additional purple shaded region marks the domain where the $\l_5$ interaction is not in equilibrium at $T_{\rm dec}^{\rm inv}$, although actual decoupling of $B-L$ violation happens at $T_{\rm dec}^{\rm sc }\ll T_{\rm dec}^{\rm inv}$ for a larger $M_1$. In that case,  the same spectator coefficient applies.

\begin{figure*}[t]
	\includegraphics[width=1\textwidth]{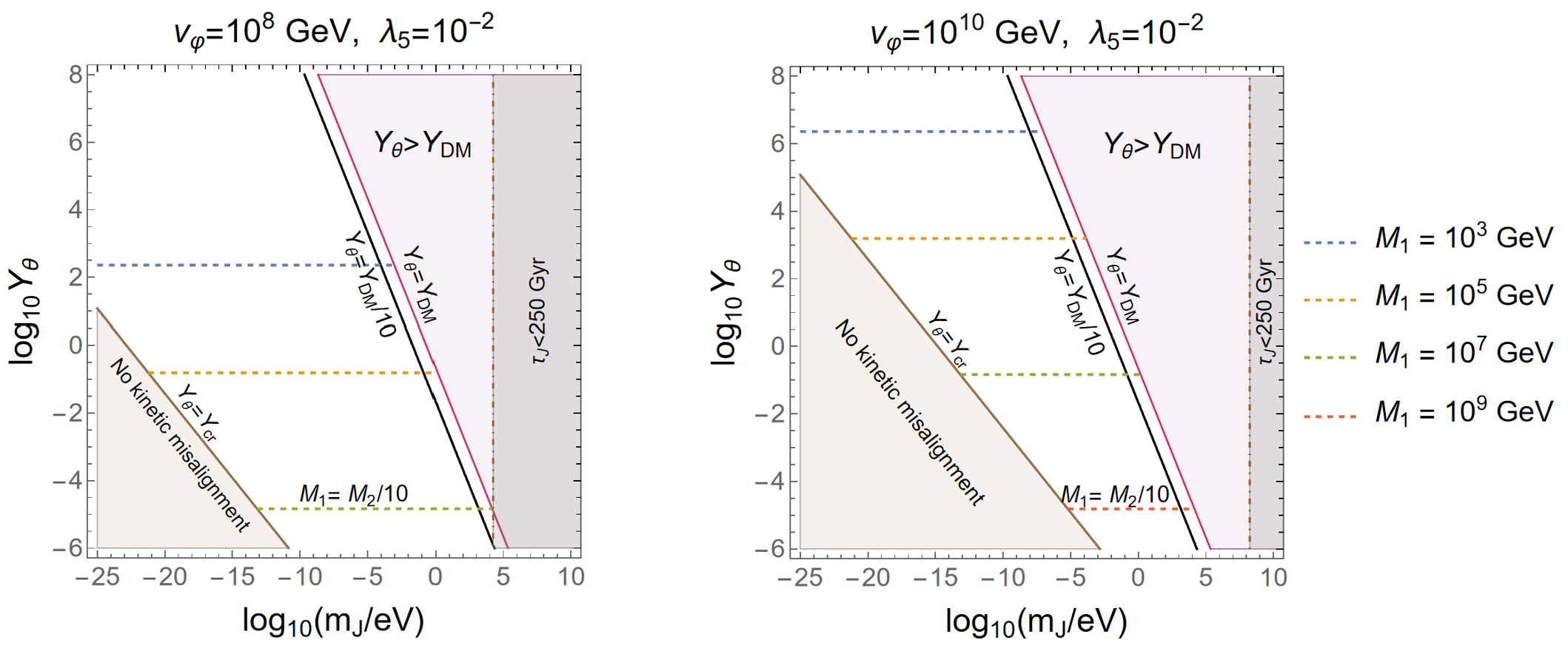}
	\caption{Parameter space in $Y_\theta$ vs $m_J$. The dashed horizontal lines reproduce the observed baryon asymmetry. The brown curve indicates the kinetic-misalignment threshold $Y_\theta=Y_{\rm cr}$. The pink shaded regions are excluded by the Majoron overproduction. In the brown shaded region, the kinetic misalignment is ineffective.}
	\label{fig:mj}
\end{figure*}

It is perhaps pertinent to point out here that if the $\l_5$ interaction reaches equilibrium before the electroweak scale, a fraction of $B-L$ asymmetry generated from spontaneous leptogenesis is distributed to the IHD sector, leading to a nonvanishing inert chemical potential $\mu_\eta\neq 0$. It is thus important to check that such inert asymmetry does not overproduce the DM abundance. Interestingly, the phenomenologically relevant region naturally leads to two distinct spaces.  
First, if $\ei$ contributes $100\%$ to DM, the bound on the inelastic DM-nucleus scattering requires a neutral inert scalar mass splitting corresponding to $\l_5\gtrsim 10^{-6}$ for sub-TeV DM. In this regime $\l_5$ interaction equilibrates before the electroweak symmetry breaking scale and remains efficient below $T_{\rm sp}$. Then, once electroweak symmetry is broken and SM chirality changing interactions become effective, the Higgs chemical potential becomes strongly suppressed, and the relation $\mu_\eta=\mu_H$ drives the IHD asymmetry to a negligible value. In particular, for $m_{\eta_I}\simeq553~{\rm GeV}$, the $\l_5$ interaction decouples only after Boltzmann suppression of the inert scalars becomes important. The representative decoupling temperatures are $T_{\l_5}^{d}\simeq35~{\rm GeV}$ for $\l_5=10^{-4}$ and $T_{\lambda_5}^{d}\simeq20~{\rm GeV}$ for $\l_5=10^{-2}$. These temperatures lie below $T_{\rm sp}$. 
On the other hand, if $\ei$ remains subdominant while the Majoron provides the dominant contribution,  smaller values of $\l_5$ can be allowed. For sufficiently small $\l_5$, the corresponding interaction never reaches thermal equilibrium. In this case, the inert sector does not participate in the redistribution of the $B-L$ asymmetry, and hence no $\mu_\eta$ is induced.

Next, we examine the possible Majoron relic abundance induced by the angular motion of the Majoron. As already discussed in Sec.~\ref{sec:lepto2}, the kinetic misalignment condition is controlled by $v_\vp$ and the mass of the Majoron $m_J$ (see Eq~\eqref{eq:ycr}). Remember also that the yield of Majoron has to satisfy $Y_{\rm cr}<Y_\theta \leq Y_\theta^{\rm DM}$, so its contribution to the DM relic does not overclose the Universe in the regime of the kinetic misalignment. With these conditions in mind, we now show in Fig.~\ref{fig:mj} the allowed and excluded regions in the $Y_\theta$ vs $m_J$ plane for a Majoron having a specific cosine potential given in Eq.~\eqref{eq:cosine}. The two panels here correspond to $v_\vp=10^8$ and $10^{10}$ GeV respectively with fixed $\l_5=10^{-2}$.

As shown in both panels, 
the pink solid contours denote the Majoron yield required to reproduce observed DM abundance (assuming $\ei$ contribution to DM is negligible), i.e.,  $Y_\theta= Y_\theta^{\rm DM}$, while the pink shaded region, where $Y_\theta>Y_{\theta}^{\rm DM}$ holds, is excluded as Majoron abundance here would overclose the Universe. 
On the other side, the solid brown contours show the critical yield $Y_{\rm cr}$ required for efficient kinetic misalignment. Thus, the unshaded region between $Y_\theta=Y_{\rm cr}$ and $Y_\theta=Y_\theta^{\rm DM}$ contours is the phenomenologically relevant region where a multicomponent DM scenario can be realized. For example, black contours inside this region indicate the scenario where Majoron DM contributes $10\%$ of the observed DM relic. The remaining $90\%$ of the DM relic then has to be explained by $\ei$.

In addition to reproducing the observed relic abundance, a viable Majoron DM component must be sufficiently long lived on cosmological timescales. For $m_J>2m_{\nu_i} (i=1,2,3)$, where $m_i$ are neutrino masses as described in Sec.~\ref{sec:neutrino}, the Majoron can decay into pair of neutrinos.
The gray shaded regions in Fig.~\ref{fig:mj} correspond to $m_J>10^5$ eV (left) and $m_J>10^8\,{\rm eV}$ (right), which are excluded by the lifetime bound of the Majoron. The Majoron couplings to neutrinos are suppressed by the symmetry breaking scale as  $i g_{J\nu_i\nu_i} J \overline{\nu_i}\gamma^5\nu_i$ with $g_{J\nu_i\nu_i} = \frac{m_{\nu_i}}{v_\vp}$, so the decay width takes the form
\begin{align} 
	\Gamma(J\to\nu_i\nu_i)
	=
	\frac{m_J}{16\pi}
	\left(\frac{m_{\nu_i}}{v_\vp}\right)^2 \sqrt{1-\frac{4 m_{\nu_i}^2}{m_J^2}} ,\quad\text{for}~ m_J>2m_{\nu_i}.
\end{align}
Requiring the Majoron to be cosmologically stable, i.e., $\tau_J=\Gamma_J^{-1} > t_{\rm Universe}\simeq 13.8$ Gyr, gives an upper bound on the allowed decay rate.  A more stringent bound on Majoron decay rate $\tau_J=\Gamma_J^{-1} > 250$ Gyr comes from combining CMB and large-scale structure data for decaying  DM~\cite{Simon:2022ftd}. Equivalently, an upper bound on the combination $m_J/v_\vp^2$ (as indicated by the gray shaded region in Fig.~\ref{fig:mj}) can be constructed.
For larger $v_\vp$, the Majoron-neutrino coupling becomes weaker and the decay is more suppressed. A heavier Majoron is then allowed while still satisfying $\tau_J>250$ Gyr. This is why the dot-dashed vertical bound from $J\to\nu_i\nu_i$ shifts to larger $m_J$ as $v_\vp$ is increased. 
For TeV scale RHNs, the most stringent constraint on $m_J$ is obtained for the lowest possible value of $v_\vp=10^4~{\rm GeV}$ as required to maintain the assumed RHN mass hierarchy, yielding approximately $m_J\lesssim0.0182$ eV.

We note that in the minimal scotogenic model with two RHNs, the same bound exists for the Majoron mass satisfying $m_J>2 m_2 \simeq 0.017$ eV. This is due to the fact that the lightest neutrino is massless, $m_1=0$ for a minimal normal ordered light neutrino system leading to a vanishing Majoron-$\nu_1$ coupling.
Thus, with $m_J\lesssim0.017$ eV, the Majoron becomes kinematically forbidden to decay to neutrinos, making it cosmologically stable against neutrino pair decays.
Combining the relic-density requirement, the condition for efficient kinetic misalignment, and the cosmological stability bound, we then find that the sub-eV Majoron region remains consistent with both the successful baryogenesis and the observed DM relic abundance for all the allowed values of the symmetry breaking scale, $v_\vp>10^4$ GeV.

Finally, we comment on the possible contribution of thermal Majoron quanta to dark radiation during the BBN epoch. Scattering processes such as $\ell_\alpha\eta\to JN_i$, $\eta N_i\to J\ell_\alpha$, $\ell_\alpha N_i\to J\eta$, and $N_iN_i\to JJ$ can, in principle, thermalize the Majoron  in the early Universe. However, once the temperature drops below the RHN masses, these interactions rapidly become inefficient due to the Boltzmann suppression of the RHN number density. Consequently, the Majoron also decouples from the thermal bath at approximately the same temperature. For sub-TeV RHNs, this decoupling temperature is typically $T_J=\mathcal{O}(100)$ GeV, implying that the Majoron having sub-eV masses remains highly relativistic at freeze-out. After decoupling, the Majoron temperature redshifts as $T_J\propto a^{-1}$. Then, the Majoron contribution to the radiation energy density at neutrino decoupling can be expressed as
\begin{align}
	\Delta N_{\rm eff}
	=\frac{4}{7}
	\left(\frac{T_J^{\rm BBN}}{T_\nu^{\rm BBN}}\right)^4=\frac{4}{7} \bigg(\frac{g_{*s}(T_{\rm MeV})}{g_{*s}(T_J)}\bigg)^{4/3}
	\simeq 0.025,
\end{align}
which is well below the current observational bounds on $\Delta N_{\rm eff}$~\cite{Planck:2018vyg} but within the target sensitivity of future CMB probes~\cite{Trendafilova:2026xtu}. Thus, the sub-eV Majoron should not affect the history of the Universe during BBN and CMB recombination.
Here, we used $g_{*s}(T_{\rm MeV})=10.75$ and $g_{*s}(T_J)=110.75$ including one IHD in addition to the SM.

\section{Conclusion}
\label{sec:con}
To conclude, in this work, we have investigated spontaneous leptogenesis in the dynamical minimal scotogenic setup with two hierarchical RHNs, where spontaneous breaking of  a global $U(1)_{B-L}$ symmetry naturally gives rise to RHN mass and a rolling Majoron background. Through the derivative interaction
$\partial_\mu\theta\, j^\mu_{B-L}$, this Majoron, while rolling, induces an effective
chemical potential which biases the $B-L$-violating decays and inverse decays of
the lightest RHN. This provides a qualitatively different route to leptogenesis
from the conventional thermal mechanism based on CP-asymmetric RHN decays.

For the minimal hierarchical two RHN scotogenic scenario, conventional thermal leptogenesis is strongly restricted by the neutrino mass and DM constraints. The decay and inverse decay of the lightest RHN always in this case lie in the strong washout regime, rendering the conventional thermal setup ineffective for $M_{1}\lesssim 10^{11}$ GeV. In contrast, the same 
strong washout behavior becomes advantageous in the spontaneous leptogenesis setup as it can efficiently bias the energy of the particle and antiparticle. Utilizing this fact, we have shown that this mechanism can reproduce the observed baryon asymmetry for hierarchical RHNs with
$M_1\gtrsim m_{\eta}$ as long as RHN decay and inverse decay to lepton and inert scalar remain efficient, without requiring a quasi-degenerate spectrum or additional RHNs.

The viability of this low-scale realization is tied to the DM and neutrino mass phenomenology of the inert sector as well as the Majoron dynamics. We have taken
$m_{\ei}$, $\l_{\rm DM}$, $\l_3$, and $\l_5$ as the independent parameters relevant for the IHD dynamics. The relic density requirement, DD limits, LEP constraints, electroweak precision
data, and vacuum stability conditions select the heavy CP-odd $\ei$ region, i.e., $m_{\ei}>400$ GeV. In particular, the viable WIMP regime, when considered as the dominant DM component, begins around
$m_{\ei}\simeq 550\,{\rm GeV}$ and extends to the TeV scale for the range of
quartic couplings considered in this work. The same parameter region is also
consistent with the nonzero mass splitting required to suppress dangerous
inelastic $Z$-mediated DD. We further observe that realizing the observed baryon asymmetry with the sub-TeV RHNs satisfying the condition $M_1>m_{\ei}$ requires a sufficiently large initial Majoron angular yield as long as symmetry breaking scale is high. Reducing symmetry breaking scale can relax this observation. 

Another interesting outcome of our analysis is the role of $\l_5$ mediated interactions. We have found that 
besides controlling the $\er-\ei$ mass splitting
and entering in the radiative neutrino mass matrix, 
$\l_5$ controls the interaction $(\eta^\dagger \Phi)^2$, which not only determines whether a DM asymmetry survives, but also influences the generation of the final baryon asymmetry. When these $\l_5$ mediated processes
are active near the RHN inverse decay decoupling, they modify the spectator relations, thereby affecting the conversion between the Majoron-induced charge density and the final baryon asymmetry. On the other hand, if they remain efficient through the electroweak symmetry breaking epoch, they also erase any would-be inert asymmetry.
We have found that within the phenomenologically allowed
$\lambda_5$ range where inert scalar is the dominant DM, its abundance is dominated by the symmetric $\ei$ component rather than by an asymmetric relic. On the other hand, when Majoron dominates the DM relic, the value of the lower bound of $\l_5$  can be reduced. In that case, $\l_5$ mediated interactions can never attain equilibrium. As a result, it does not play any role in spectator processes for spontaneous leptogenesis.

We have also identified the cosmological consistency conditions associated with the rolling Majoron background. The radiation-dominated treatment adopted in this work requires the Majoron kinetic energy to remain subdominant at the time of baryon asymmetry freeze-out. This condition restricts the allowed combinations of $Y_\theta$, $v_\vp$, and $M_1$ parameter regions in which the Majoron kinetic energy dominates and leads to a kination-like expansion history, which requires a separate analysis. 

Finally, the coherent Majoron population can also contribute to the DM density without
spoiling the successful generation of the baryon asymmetry in this setup. The sub-eV Majoron condensate is found to simultaneously remain cosmologically stable, avoid overclosing the Universe, and support the rolling background required for successful spontaneous leptogenesis, even when the RHNs lie near the TeV scale. The stability bound on Majoron mass becomes weaker for larger $v_\varphi$, allowing heavier Majorons to remain long-lived on cosmological time scales. Moreover, thermal Majoron quanta with sub-eV
masses decouple early (even for sub-TeV RHNs) and give only a small dark-radiation contribution,
$\Delta N_{\rm eff}\simeq 0.025$, which is below the present limits but can be probed by future CMB measurements.

Our results therefore, show that the dynamical scotogenic model provides an
economical realization of spontaneous leptogenesis in which the origin of neutrino masses, IHD DM, and the baryon asymmetry are linked to the same symmetry breaking structure. The possibility of successful leptogenesis with TeV-scale hierarchical RHNs makes this framework a unique and testable
alternative to high-scale thermal leptogenesis, with correlated probes from
DD experiments, inert scalar and missing energy searches at
colliders, and future measurements of dark radiation.

\section*{Acknowledgment}
We thank Prof. Eung Jin Chun for discussion at the initial stage of the project.
This research was supported by the Chung-Ang University research grant in 2026.
The work is also supported in part by Basic Science Research Program through the National
Research Foundation of Korea (NRF) funded by the Ministry of Education, Science and
Technology (NRF-2022R1A2C2003567).

\section*{Appendix}
In the appendix, we derive the baryon asymmetry conversion factor, namely, $c_B$, in the scotogenic model, depending on the decoupling temperature for the $B-L$ violating process mediated by $N_1$, $T_{\text{dec}}$ for large $\l_5$ values. \\

\noindent
\underline{1. The case with  $T_{\text{dec}}< 10^{5}$ GeV}: \\
First, at a low decoupling temperature satisfying $T_{\text{dec}}< 10^{5}$ GeV, there is no flavor dependence \cite{Domcke:2020kcp}, so we can take 
\begin{align}
	\mu_{ui}\equiv& \mu_{u}=\mu_{c}=\mu_{t}\notag\\
	\mu_{di}\equiv& \mu_{d}=\mu_{s}=\mu_{b}.\notag
\end{align}
Then, we can simply write following chemical equilibrium relations (with $\mu_N=0$)
\begin{align}
	\mu_{L}+3\mu_{Q}=& \ 0\notag\\
	\mu_{e}-\mu_{L}=& \ \mu_{H}\notag\\
	\mu_{L}+\mu_{\eta}=& \ \frac{\Dot{\theta}}{2}\notag\\
	\mu_{ui}-\mu_{Q}=& \ \mu_{H}\notag\\
	\mu_{di}-\mu_{Q}=& \ -\mu_{H}\notag\\
    \mu_H=& \ \mu_{\eta}.
\end{align}
So, we have 7 chemical potentials, $\mu=(\mu_e,\mu_L,\mu_u,\mu_d,\mu_Q,\mu_H,\mu_{\eta})^T$
and  the hypercharge neutrality condition is given by
\begin{align}
	\sum_{i=1}^7g_i n^Y_i \mu_i=0
\end{align}
where $n^Y=(-2,-1,\frac43,-\frac23,\frac13,1,1)^T$ is the hypercharge vector and $g=(3,6,9,9,18,4,4)^T$ is the vector for degrees of freedom.
As a result, we obtain $c_B$ from the following relation:
\begin{align}
	\mu_B=\sum_{i=1}^7 g_i n^B_i \mu_i \equiv c_B \Dot{\theta}
\end{align}
where $n^B=(0,0,\frac13,\frac13,\frac13,0,0)$.
Thus we get $c_B=-4/3$. \\

\noindent
\underline{2. The case with  $4.5 \times 10^6 \text{ GeV}\lesssim T_{\text{dec}}\lesssim 1.1\times 10^9 \text{ GeV}$}: \\
For $4.5 \times 10^6 \text{ GeV}\lesssim T_{\text{dec}}\lesssim 1.1\times 10^9 \text{ GeV}$, we need to carefully consider the flavor dependence because the up-type and down-type Yukawa couplings for the 1st generation and the electron Yukawa coupling are not efficient any more. In this case, we take the chemical potential vector, $\mu=\left(\{\mu_{e_{Ri}}\}, \{ \mu_{L_i} \}, \{ \mu_{u_{Ri}} \}, \{ \mu_{d_{Ri}} \}, \{ \mu_{Q_i} \},\mu_{H},\mu_{\eta}\right)^T$ where $i=1,2,3$ are flavor indices, so there are 17 total number of chemical potentials.

For electroweak and strong sphaleron processes,, there are two chemical equilibrium relations,
\begin{align}
    \sum^3_{i=1}\left(\mu_{Li}+3\mu_{Qi}\right)=&0\notag\\
    \sum^3_{i=1}\left[2\mu_{Q_i}+(\mu_{u_{Ri}}+\mu_{d_{Ri}})\right]=&0
\end{align}
For the up-type Yukawa interactions,
\begin{align}
   -\mu_{u_{Ri}}+\mu_{Q_i}+\mu_H=&0
\end{align}
where $i=2,3$.

For the diagonalized Yukawa couplings for up-type quarks, the Yukawa couplings for down-type quarks can't be diagonalized before electroweak symmetry breaking. Thus, we also need to take into account the flavor mixing,
\begin{align}
    -\mu_{d_{Ri}}+\mu_{Q_i}+\mu_H=&0\notag\\
    -\mu_{d_{R2}}+\mu_{Q_3}+\mu_H=&0
\end{align}
where $i=2,3$.

We are also considering the Yukawa interactions for neutrinos. In the strong washout region where $K \gg 4$, we have 
\begin{align}
    \mu_{L_i}+\mu_{\eta}=&\frac{\Dot{\theta}}{2}
\end{align}
where $i=1,2,3$. We also have $\mu_{H}=\mu_{\eta}$ for a sizable value, $\lambda_5 \sim 10^{-2}$.
The above chemical interactions conserve the following three global symmetries \cite{Domcke:2020quw}:
\begin{align}
    U(1)_{e_{R1}}, \quad U(1)_{u_{R1}-d_{R1}}, \quad U(1)_{2B_1-B_2-B_3}
\end{align}
where $B_i$ is the baryon charge for flavor $i$. By assuming that the initial asymmetry corresponding to each of three charges is zero at the beginning of the Universe, we have 3 chemical potential relations:
\begin{align}
    &\mu_{e_{R1}}=0\notag\\
    &\mu_{u_{R1}}-\mu_{d_{R1}}=0\notag\\
    &2(\mu_{u_{R1}}+\mu_{d_{R1}}+2\mu_{Q_1})-\sum^3_{i=2}(\mu_{u_{Ri}}+\mu_{d_{Ri}}+2\mu_{Q_i})=0.
\end{align}
We also include the hypercharge neutrality condition, 
\begin{align}
	\sum_{j=1}^{17}g_j n^Y_j \mu_j=0
\end{align}
where the hypercharge vector and the vector for degrees of freedom are given by
\begin{align}
n^Y=&(-2,-2,-2,-1,-1,-1,\frac43,\frac43,\frac43,-\frac23,-\frac23,-\frac23,-\frac13,\frac13,\frac13,1,1)^T, \\ g=&(1,1,1,2,2,2,3,3,3,3,3,3,6,6,6,4,4)^T,
\end{align}
respectively.

Now we have 17 chemical potential relations for unknown 17 chemical potentials , so we can obtain $c_{B-L}$ from the following relation:
\begin{align}
	\mu_{B-L}=\sum_{i=1}^{17} g_i n^{B-L}_i \mu_i \equiv c_{B-L} \Dot{\theta}
\end{align}
where $n^{B-L}=(-1,-1,-1,-1,-1,-1,\frac13,\frac13,\frac13,\frac13,\frac13,\frac13,\frac13,\frac13,\frac13,0,0)$. Thus we get $c_{B-L}=-\frac{65}{19}$.

The analysis can be straightforwardly extended to smaller $\l_5$ even when $T_{\rm dec}$ remains above the equilibration temperature of the $\l_5$ mediated interaction. In this regime, however, $\mu_\eta$ must be treated as an independent chemical potential in addition to the Majoron-induced chemical-potential bias. For the low-scale leptogenesis scenario in the absence of any additional mechanism capable of generating a sizable primordial $\eta$ asymmetry, one can take $\mu_\eta=0$ for small $\l_5$ values without loss of generality.

Next, we derive the sphaleron conversion factor $c_{\text{sph}}$ which is defined by $\mu_{B}=c_{\text{sph}}\mu_{B-L}$.
After the decoupling of the $B-L$ violating interaction at $4.5\times 10^6 \text{ GeV}\lesssim T_{\text{dec}}\lesssim 1.1\times 10^9 \text{ GeV}$ and a non-zero $\mu_{B-L}$ is fixed, the electroweak sphaleron process is efficient until $T_{\text{EW}}\simeq 130 \text{ GeV}$. Around $T\gtrsim T_{\text{EW}}$, all the Yukawa chemical interactions are efficient, so there is no flavor dependence while all the chemical potential relations are rearranged.
For given 7 chemical potential, $\mu=(\mu_e,\mu_L,\mu_u,\mu_d,\mu_Q,\mu_H,\mu_{\eta})^T$, we have
\begin{align}
	\mu_{L}+3\mu_{Q}=& \ 0\notag\\
	\mu_{e}-\mu_{L}=& \ \mu_{H}\notag\\
	\mu_{ui}-\mu_{Q}=& \ \mu_{H}\notag\\
	\mu_{di}-\mu_{Q}=& \ -\mu_{H}\notag\\
    \mu_H=& \ \mu_{\eta}
\end{align}
We also have the hypercharge neutrality condition and the condition for conserving the $\text{B-L}$ asymmetry,
\begin{align}
	\sum_{i=1}^7g_i n^Y_i \mu_i=& \ 0\notag\\
    \sum_{i=1}^7g_i n^{\text{B-L}}_i \mu_i=& \ \mu_{B-L}=c_{B-L}\Dot{\theta}
\end{align}
where $n^Y=(-2,-1,\frac43,-\frac23,\frac13,1,1)^T$ is the hypercharge vector, $n^{\text{B-L}}=(-1,-1,\frac13,\frac13,\frac13,0,0)^T$ and $g=(3,6,9,9,18,4,4)^T$ is the vector for degrees of freedom. We also recall $\mu_{B-L}=c_{B-L}\Dot{\theta}=-\frac{65}{19}\Dot{\theta}$.

We can obtain $c_B$ from the following relation:
\begin{align}
	\mu_B=\sum_{i=1}^7 g_i n^B_i \mu_i \equiv c_B \Dot{\theta}
\end{align}
As a result, we get $c_B=-\frac{520}{437}$ and the conversion factor is $c_{\text{sph}}\equiv \frac{c_{\text{B}}}{c_{B-L}}=\frac{8}{23}$.


\noindent
\underline{3. The case with $10^5$ GeV $\lesssim T_{\text{dec}} \lesssim 4.5\times 10^6$ GeV}: \\
Lastly, for $10^5$ GeV $\lesssim T_{\text{dec}} \lesssim 10^6$ GeV, only the electron Yukawa coupling is decoupled. For $10^6$ GeV $\lesssim T_{\text{dec}} \lesssim 4.5\times 10^{6}$ GeV, the up-quark Yukawa and electron Yukawa couplings are decoupled. But the two cases give rise to the same results, $\mu_{B-L}=-\frac{83}{22}\Dot{\theta}$, because the strong sphaleron interaction basis can be superposed by all the Yukawa interaction basis which also includes up-quark Yukawa interactions.

\bibliography{LL_ref.bib}
\end{document}